\documentclass{jfm}
\usepackage{graphicx}
\usepackage{epstopdf, epsfig}
\usepackage{tikz}
\usepackage{color}

\shortauthor{Michael Eigenbrod and Steffen Hardt}

\title{The effective shear and dilatational viscosity of a particle-laden interface in the dilute limit}

\author{Michael Eigenbrod\aff{}
 \and Steffen Hardt\aff{} \corresp{\email{hardt@nmf.tu-darmstadt.de}}}

\affiliation{\aff{}{\color{black}Technische Universit\"{a}t Darmstadt, Department of Mechanical Engineering, Institute for Nano- and Microfluidics, Germany}}

\begin{document}

\maketitle
\begin{abstract}
The effective dilatational and shear viscosities of a particle-laden fluid interface are computed in the dilute limit under the assumption of an asymptotically vanishing viscosity ratio between both fluids. Spherical particles with a given contact angle of the fluid interface at the particle surface are considered. A planar fluid interface and a small Reynolds number are assumed. The theoretical analysis is based on a domain perturbation expansion in the deviation of the contact angle from $90^\circ$ up to the second order. The resulting effective dilatational viscosity shows a stronger dependence on the contact angle than the effective shear viscosity, and its magnitude is larger for all contact angles. As an application of the theory, the stability of a liquid cylinder decorated with particles is considered. The limits of validity of the theory and possible applications in terms of numerical simulations of particle-laden interfaces are discussed.    
\end{abstract}


\section{Introduction}\label{sec:s1}
Suspensions of particles in a bulk fluid commonly occur in nature and in several engineering applications. \cite{Einstein1906a} was the first to quantitatively determine the viscosity of a dilute suspension of spherical particles. More specifically, the presence of the particles leads to an increase in the energy dissipation inside the fluid which is effectively related to an increase in viscosity. The basis of these calculations was to compute the rate of viscous dissipation in a large control volume concentric with a particle. Another milestone in suspension rheology is the work of \cite{Batchelor1970}. By employing volume averaging of the stresses inside the suspension, he found an elegant way of calculating the effective suspension properties, which in its most general form applies to arbitrary particle concentrations and shapes. His analytical solutions, however, are limited to dilute suspensions where only a single particle or the interaction of two particles are taken into account. \cite{Batchelor1972} have managed to go beyond that limit and computed the viscosity of suspensions up to second order in the volume concentration. Due to the hydrodynamic interactions of sufficiently small particles at higher concentrations, Brownian motion has to be taken into account and can be represented as a stochastic force \citep{Batchelor1976,Batchelor1977}. However, in most of the work reported in the literature, higher-order approximations (in the concentration) are typically achieved through numerical simulations \citep{Brady2000,Sierou2008}. \\

Whenever particles are attached to fluid interfaces, i.e. interfaces between a gas and a liquid or between two immiscible liquids, additional physical effects, such as capillarity, superpose to the effects being present in the bulk of a fluid. In the past decades, systems with particles attached to fluid interfaces have been studied quite intensely, since the particles bring along unique properties not found for bare interfaces. Interfacial particles mechanically stabilize droplets \citep[e.g.,][]{Binks2002,Aveyard2003,Dickinson2010,Wu2016,DeCorato2018}. These composite structures are often referred to as liquid marbles \citep{Aussillous2001}. The evaporation rate of liquid marbles on surfaces is lower than that of bare droplets, since no liquid is in direct contact with the substrate \citep{McHale2011}. Liquid marbles enable the creation of hollow granules \citep{Eshtiaghi2010}, are usually more mobile on surfaces than bare droplets and can be positioned and transported via gravity (more precisely than a bare droplet since contact angle hysteresis is negligible) \citep{Aussillous2001}, applied magnetic \citep{Bormashenko2008} or electric fields \citep{Newton2007}. Owing to these properties, potential applications are abundant. Liquid marbles may be used in pharmaceutical, medical, chemical, and cosmetic industries \citep{Avramescu2018}. Apart from that, particles attached to fluid interfaces play a role in creating nanostructured materials based on the Langmuir and Langmuir-Blodgett techniques \citep{Park2011a} or in the stabilization of emulsions \citep{Wu2016}.\\

Potentially inspired by the manifold applications of interfacial particles, \cite{Lishchuk2009} were the first to calculate the effective dilatational and shear viscosity of a particle-laden interface in the dilute limit (surface concentration $\phi_L \ll 1$) for particles having a contact angle of $\alpha = 90^\circ$. They used a method similar to that of \cite{Einstein1906a}. They further verified their results with Lattice-Boltzmann simulations and demonstrated good agreement if $\phi_L \le 0.15$. Later on, \cite{Lishchuk2014} modeled the effective dilatational viscosity of an interface densely decorated with particles, where Brownian motion was neglected and the contact angle was fixed to $\alpha = 90^\circ$.  By employing a toroidal coordinate system, \cite{Lishchuk2016} calculated the effective dilatational surface viscosity in the dilute limit as a function of the contact angle $\alpha$, when both phases have a large viscosity contrast. The resulting integral had to be solved numerically.  \\

What has not been achieved up to now is to determine closed-form analytical expressions for the dilatational and shear viscosity of a particle-laden interface, valid for a broad range of contact angles $\alpha$. {\color{black}Although a semi-analytical solution for the effective dilatational viscosity has already been reported, the dependence of the interfacial shear viscosity on the contact angle on the particle surface has remained obscure, limiting the practical use of prior studies significantly. The main objective of the present work is to close the gap previous studies have left and to shed more light in the rheology of interfacial suspensions}. In section \ref{sec:Prob}, we describe the problem, the model assumptions and the governing equations together with the boundary conditions. {\color{black} The stress averaging method of \cite{Batchelor1970} is applied to interfacial suspensions.} In section \ref{sec:Eff}, we make use of this method to calculate the effective surface viscosities up to second order in $\beta = \cos(\alpha)$. In section \ref{sec:Diss}, we discuss our results, apply them to the stability problem of a free liquid jet and describe the application of our theory in terms of numerical simulations of particle-laden interfaces. Finally, in section \ref{sec:Conc} we summarize the main results of our work and give an outlook to further studies. 

\section{Problem formulation and model assumptions}\label{sec:Prob}
\begin{figure}
  \centerline{\includegraphics[width=0.95\textwidth]{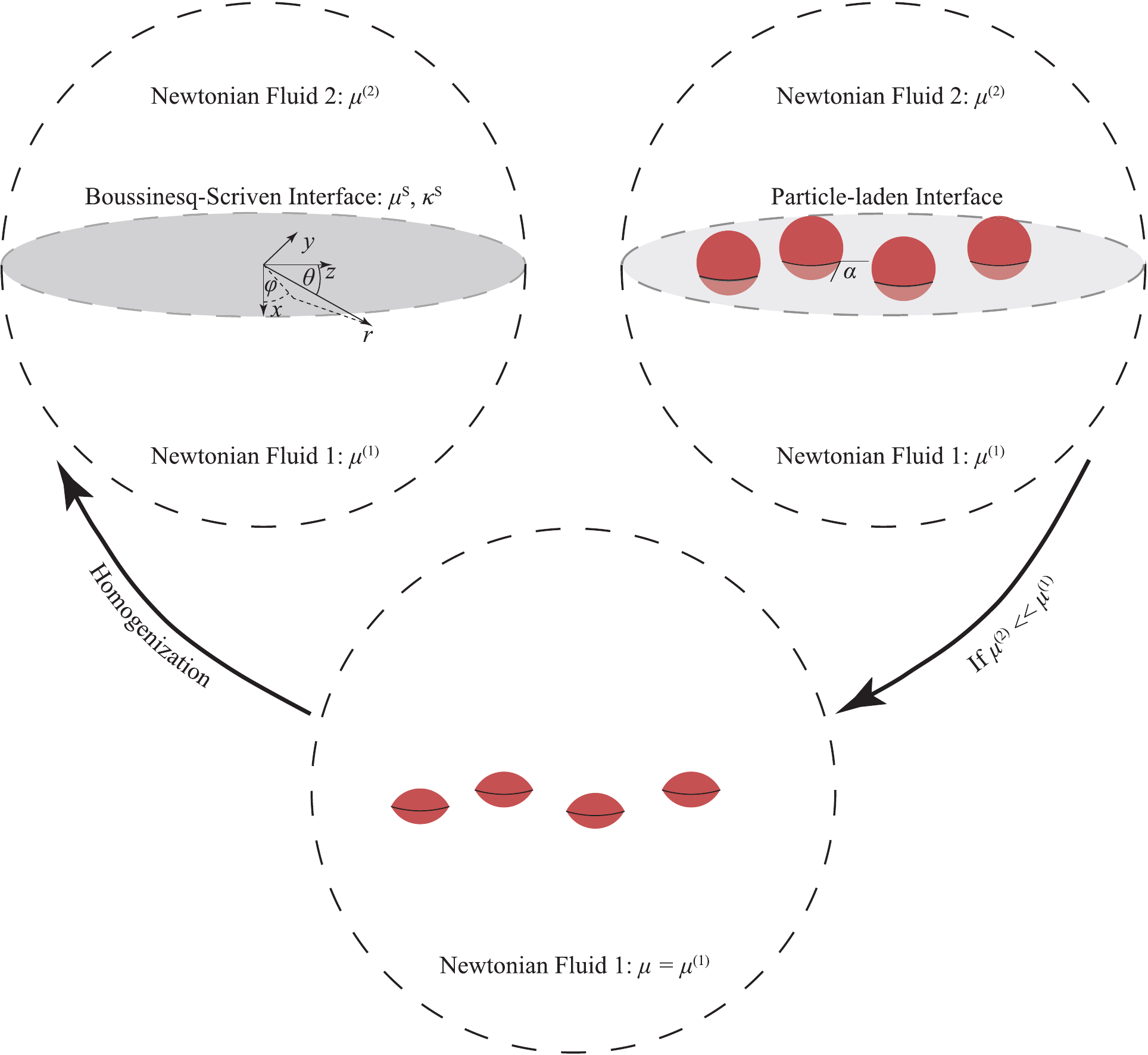}}
  \caption{{\color{black}Visualization} of the similarities used for the computation of the effective surface viscosities represented by a Boussinesq-Scriven constitutive law (top left), see equation (\ref{eq:Prob_14}). The basis of the calculation are spherical particles attached to an ideal interface (massless and vanishing stresses) which is equivalent to mirror-fused particles suspended in an infinite bulk fluid if the viscosity ratio $\mu_2 / \mu_1$ vanishes asymptotically.}
\label{fig:f1}
\end{figure}

We consider a particle-laden fluid interface, consisting of identical spherical particles having a radius $a$ and a contact angle $\alpha$. In the absence of particles, the infinitesimally thin interface [so-called dividing surface or Gibbs interface \citep{Slattery2007}] separating both phases is assumed to be \textit{ideal} [massless/incompressible with vanishing surface stresses \citep{Wang2011}]. Ideal interfaces are commonly assumed in the absence of surfactants or impurities \citep[e.g.,][]{Brenner1991}. We always consider an ideal interface and aim at computing the interfacial viscosities arising when we add particles. We do this in the famework of a homogenization method. That is, the particle-laden interface is represented by a homogenenous interface having effective viscosities (figure \ref{fig:f1}, top left and top right). 
 
In general, hydrodynamic interparticle interactions, Brownian motion as well as long-range capillary interactions (due to a deformation of the fluid interface) bring along many complications that leave us with a very complex problem that is usually inaccessible to analytical methods. To keep the problem tractable, the following assumptions are made:\\
 
\begin{enumerate}
\item Both fluid phases are Newtonian and incompressible; \label{itm:1}
\item The Reynolds number in phase $1$ is small, i.e., $\Rey^{(1)} \ll 1$;\label{itm:2}
\item The radius of curvature of the fluid interface separating both phases is much larger than the characteristic size $a$ of the particles {\color{black}and the capillary number $\mathrm{Ca}$ is sufficiently small};\label{itm:3}
\item The viscosity ratio between both phases vanishes asymptotically, i.e., $\mu^{(2)}/\mu^{(1)}\to 0$;\label{itm:4}
\item The interfacial particles are neutrally buoyant, rigid, of spherical shape with radius $a$, while the fluid interface assumes a contact angle of $\alpha$ on the particle surface; \label{itm:5}
\item The interfacial suspension is dilute; \label{itm:6}
\item Brownian motion is negligible;\label{itm:7}
\item {\color{black}No external force or torque is applied to the particles and the Stokes numbers reflecting the relaxation time scales for the linear and angular momentum of the particles are sufficiently small.}\label{itm:8}\\
\end{enumerate}
Assumption (\ref{itm:2}) simplifies the momentum balance by neglecting any transient and convective terms. {\color{black}For small capillary numbers, the stress balance at the interface can be expanded into a perturbation series in $\mathrm{Ca}$. }As long as the radius of curvature of the interface is much larger than the characteristic size of a particle [assumption (\ref{itm:3})], the curvature of the interface can be neglected {\color{black}in the zeroth order in $\mathrm{Ca}$}. Assumption (\ref{itm:4}) is a reasonable approximation for any gas-liquid system, since, taking the air-water system as an example, we obtain $\mu^{\mathrm{air}} / \mu^{\mathrm{water}} \approx 0.02$ \citep{Petkov1995}. When assuming that the spherical particles are neutrally buoyant and rigid, the fluid interface remains flat even if particles are attached to it [assumption (\ref{itm:5})]. Generally, fluid interfaces are free of shear stresses (if no Marangoni stresses are present) and together with assumption (\ref{itm:3}), the interface therefore has all properties of a symmetry plane. Consequently, the velocity and pressure field around the particles can be calculated by replacing the interfacial particles with particles consisting of two fused mirror-reflected spherical caps in an unbounded fluid, as shown in figure \ref{fig:f1} (top right and bottom). This kind of symmetry argument has already been used to calculate the drag force acting on an interfacial particle driven by an applied force \citep{Dorr2015,Dorr2015a}, for the calculation of the electrophoretic mobility of an interfacial particle \citep{Eigenbrod2018}, as well as in the {\color{black}semi-analytical} calculation of the effective dilatational viscosity of particle-laden interfaces \citep{Lishchuk2016}. Therefore, the upper phase (phase 2) does not affect the interfacial particles. The effective interfacial properties are therefore solely determined by the stresses acting in fluid phase 1. From now on, all material properties appearing in the equations correspond to phase 1 and we suppress the superscripts.\\

Interfacial particles may interact through different mechanisms, e.g. through hydrodynamic interactions due to the mean relative velocity of the particles or through velocity fluctuations due to Brownian motion. Furthermore, it has been shown that when interfacial particles execute an oscillating motion normal to the interface \citep{Toro-Mendoza2017}, the resulting capillary waves induce an interaction. A particle-particle interaction is also induced by the dynamic interfacial deformation of particles moving along a fluid interface \citep{Dorr2015}. Even in the static case the deformation of the interface leads to capillary interactions between particles \citep{Leandri2013,Galatola2014}. Naturally, all of these interactions decay with the interparticle distance. In a similar way as in the calculation of the effective viscosity of a bulk suspension we need to assume that the interparticle distance is large enough to render these interactions negligible [assumption (\ref{itm:6})]. Therefore, among others, it is assumed that the perturbation velocity field due to the presence of a particle does not influence neighbouring particles. It should be noted that \cite{Batchelor1977} has shown that the dilute-limit solution by \cite{Einstein1906a} holds even in the presence of Brownian motion. However, this general statement can not be made in case of interfacial particles, due to the capillary waves mentioned above. When no external force or torque is applied to the particles (for example through an external electric or magnetic field) and inertial effects {\color{black}(in both the fluid and the particle motion)} are neglected, the particles follow the flow in such a way that the {\color{black}sums of the force/torque due to the translation/rotation of a particle and the force/torque induced by the translational/rotational velocity of the surrounding fluid are identically zero [assumption (\ref{itm:8})]. As an example, for a single spherical particle suspended in an unbounded Newtonian fluid and subjected to an arbitrary flow field, it is known that the translational and rotational velocity of the particle coincides with that of the applied velocity field \citep{Happel2012, Kim2013}.} \\

Before we go over to discuss the governing equations as well as the corresponding boundary conditions, we comment on the definition of the surface concentration by \cite{Lishchuk2009}, i.e.,
\begin{equation}
\phi_L = \frac{N \pi a^2}{A}. \label{eq:Prob_1}
\end{equation}
$A$ is the total area of the fluid interface, $N$ the number of particles attached to the interface and $a$ the radius of the particles. For contact angles deviating from $\alpha = 90^\circ$, however, the area fraction of the fluid interface cutting through the particle is smaller than that and given by
\begin{equation}
\phi = \frac{N \pi a^2 [1-\cos^2(\alpha)]}{A}. \label{eq:Prob_2}
\end{equation}
We later compare the {\color{black}semi-analytical} results by \cite{Lishchuk2016} to our theoretical results and therefore have to keep both definitions in mind. Apart from that, we do not use the definition from equation (\ref{eq:Prob_1}). 

\subsection{Governing equations and boundary conditions}\label{sec:Prob_1}
As already pointed out when discussing the underlying assumptions, particles at ideal interfaces can be represented by mirror-fused particles in a bulk fluid, see figure \ref{fig:f1} (top right and bottom). We therefore describe the governing equations and the corresponding boundary conditions referring to the bottom of figure \ref{fig:f1}. \\

Following assumptions (\ref{itm:1}) \& (\ref{itm:2}), the governing equations are the Stokes equations, reading
\begin{equation}
\boldsymbol{\nabla} p_\mathrm{tot} = \mu \boldsymbol{\nabla}^2 \boldsymbol{u}_\mathrm{tot}, \qquad \boldsymbol{\nabla} \cdot \boldsymbol{u}_\mathrm{tot} = 0, \label{eq:Prob_4}
\end{equation}
in which $p_\mathrm{tot}$ is the thermodynamic pressure, $\boldsymbol{u}_\mathrm{tot}$ the velocity field, and $\mu$ the dynamic viscosity (referring to phase 1). Here, $\boldsymbol{u}_\mathrm{tot} = \boldsymbol{u}^{\infty} + \boldsymbol{u}$ is the superposition of an applied velocity field $\boldsymbol{u}^{\infty}$ and a disturbance velocity field $\boldsymbol{u}$ occurring through the presence of particles. The first equation in (\ref{eq:Prob_4}) is the linearized momentum balance equation, in which the inertia terms as well as body force terms are neglected [assumption (\ref{itm:5})]. The second equation in (\ref{eq:Prob_4}) ensures incompressibility in the bulk. \\

Next we consider an arbitrary applied velocity around the particles and linearize it around the origin of the coordinate system (shown in figure \ref{fig:f1}, top left), leading to \citep{Brenner1991}
\begin{equation}
\boldsymbol{u}^{\infty}(\boldsymbol{r}) \approx \boldsymbol{U}^\infty + \boldsymbol{\Omega}^{\infty} \times \boldsymbol{r} + \boldsymbol{E} \cdot \boldsymbol{r}, \label{eq:Prob_4b}
\end{equation}
in which $\boldsymbol{r}$ is the position vector and $|\boldsymbol{r}| = r$ its length. Generally, a particle suspended in the bulk can execute a translation and rotation due to the applied velocity field $\boldsymbol{u}^{\infty}(\boldsymbol{r})$ and, if the suspension is not dilute, due to perturbations of the velocity field originating from nearby particles. The latter is ignored owing to assumption (\ref{itm:6}). Consequently, the far field boundary condition for every particle is of the same mathematical form. In the following, we represent all of these non-interacting particles by a single particle to which this boundary condition applies, located at the center of the coordinate system. The Stokes equations are linear and consequently, the boundary conditions at the surface of the representative particle as well as far away from the particle can be written as the difference between the translational and rotational velocity of the particle and the unperturbed applied, linearized velocity field leading to
\begin{equation}
\left. \boldsymbol{u}\right|_{r_\mathcal{P}} = \boldsymbol{U}^{\mathrm{eff}} + \boldsymbol{\omega}^{\mathrm{eff}} \times \frac{\boldsymbol{r}}{r} r_\mathcal{P} - \boldsymbol{E} \cdot \frac{\boldsymbol{r}}{r} r_\mathcal{P}, \mathrm{~and~} \lim\limits_{r \to \infty}  \boldsymbol{u} = \boldsymbol{0}, \label{eq:Prob_5}
\end{equation}
in which $\boldsymbol{U}^{\mathrm{eff}} {\color{black}=~}\boldsymbol{U} - \boldsymbol{U}^\infty$ and $\boldsymbol{\omega}^{\mathrm{eff}} = \boldsymbol{\omega} - \boldsymbol{\Omega}^\infty$ are the effective translational velocity, defined as the difference between the velocity of each particle and the applied translational velocity, and the effective angular velocity, respectively. $r_\mathcal{P}$ is the parametrization of the surface of the particle. The effective translational/rotational velocity is typically obtained through a force/torque balance at the particle. For spheres ($r_\mathcal{P} = a$) in an unbounded medium, it follows by Faxen's law \citep{Happel2012, Kim2013} and in the absence of external forces or torques [assumption (\ref{itm:8})] that $\boldsymbol{U}^{\mathrm{eff}} = \boldsymbol{0}$ and $\boldsymbol{\omega}^{\mathrm{eff}} = \boldsymbol{0}$, leading to the boundary condition $\left. \boldsymbol{u}\right|_{a} = - \boldsymbol{E} \cdot \frac{\boldsymbol{r}}{r} a$ at the surface of the particle. Translated to an interfacial particle, this corresponds to a particle having a contact angle of $\alpha = 90^\circ$. For non-spherical particles an effective translation and rotation might exist, which would effect the boundary conditions at the surface of the representative particle and consequently the effective viscosities. However, in appendix \ref{sec:S1} we show that the effective translational and rotational velocities do not need to be considered in the present study. Consequently, the boundary conditions at the surface of the particle correspond to a pure straining flow field.\\

At this point, we have to restrict the rate of strain tensor $\boldsymbol{E}$, since in the absence of Marangoni stresses fluid interfaces are shear-free and further, the flow is assumed to be incompressible. The rate of strain tensor then reads
\begin{equation}
\boldsymbol{E} = \left[
\begin{array}{ccc}
  -\boldsymbol{\nabla}^{\mathcal{S}} \cdot \boldsymbol{u}^{\mathcal{S}} & 0 & 0 \\  
  0 & E_{{\color{black}yy}}^\mathcal{S} & E_{{\color{black}yz}}^\mathcal{S}\\
  0 & E_{{\color{black}yz}}^\mathcal{S} & E_{{\color{black}zz}}^\mathcal{S}
\end{array}  \right], \qquad \boldsymbol{E}^\mathcal{S} = \left[
\begin{array}{cc}	
 E_{{\color{black}yy}}^\mathcal{S} & E_{{\color{black}yz}}^\mathcal{S}\\
 E_{{\color{black}yz}}^\mathcal{S} & E_{{\color{black}zz}}^\mathcal{S},
\end{array}  \right]\label{eq:Prob_6}
\end{equation}
with $\boldsymbol{\nabla}^{\mathcal{S}} \cdot \boldsymbol{u}^{\mathcal{S}} = E_{{\color{black}yy}}^\mathcal{S} + E_{{\color{black}zz}}^\mathcal{S}$. The boundary conditions for a mirror-fused bulk particle can therefore be written as
\begin{equation}
\left. \boldsymbol{u}\right|_{r_\mathcal{P}} = - \boldsymbol{E} \cdot \frac{\boldsymbol{r}}{r} r_\mathcal{P}, \mathrm{~and~} \lim\limits_{r \to \infty}  \boldsymbol{u} = \boldsymbol{0}.  \label{eq:Prob_7}
\end{equation} 
It it worth mentioning that equation (\ref{eq:Prob_7}) describes a fixed particle with a no-slip condition on its surface in pure straining flow. In summary, equations (\ref{eq:Prob_4}) \& (\ref{eq:Prob_7}) provide all necessary information to compute the flow around a mirror-fused particle in an unbounded fluid. Before we go over to present a solution strategy, we {\color{black} introduce the underlying method employed for the computation of the effective interfacial viscosities, which is closely related to the scheme presented by \cite{Batchelor1970}. Further details can be found in appendix \ref{sec:S2a}. Employing the rationale of appendix \ref{sec:S2a}, we obtain
\begin{eqnarray}
\langle \boldsymbol{\sigma}^{\mathcal{S}} \rangle : \boldsymbol{E}^\mathcal{S} &=& \frac{N}{2 A} \boldsymbol{S} : \boldsymbol{E}, \label{eq:Prob_10}\\
\boldsymbol{S} &=& \int\limits_{A_\mathcal{P}} \frac{1}{2} (\boldsymbol{r} \boldsymbol{\Pi} \cdot \boldsymbol{n} + \boldsymbol{\Pi} \boldsymbol{r}  \cdot \boldsymbol{n})-\frac{1}{3} \boldsymbol{r} \cdot \boldsymbol{\Pi} \cdot \boldsymbol{n}~\boldsymbol{I} - \mu(\boldsymbol{n} \boldsymbol{u} + \boldsymbol{u} \boldsymbol{n}) \mathrm{ d }A. \label{eq:Prob_11}
\end{eqnarray}
$\langle \boldsymbol{\sigma}^\mathcal{S}\rangle$, $\boldsymbol{\Pi}$ and $A_\mathcal{P}$ are the area-averaged deviatoric interfacial stress, the total bulk stress and the surface of a mirror-fused particle, respectively. $\boldsymbol{S}$ is the stresslet due to the particle. Making use of equation (\ref{eq:Prob_2}) we obtain
\begin{equation}
\langle \boldsymbol{\sigma}^\mathcal{S} \rangle : \boldsymbol{E}^\mathcal{S} = \frac{\phi}{2 \upi a^2 [1 - \cos^2(\alpha)]} \boldsymbol{S} : \boldsymbol{E}. \label{eq:Prob_12}
\end{equation}}
The foregoing derivation is still very general, since we have not specified a constitutive equation for the interfacial stresses. We may consider a compressible Newtonian interface, which is described by a Boussinesq-Scriven law \citep{Boussinesq1913,Scriven1960}:
\begin{equation}
\boldsymbol{\sigma}^\mathcal{S} = (\kappa^\mathcal{S} - \mu^\mathcal{S}) \left(\boldsymbol{\nabla}^\mathcal{S} \cdot \boldsymbol{u}^\mathcal{S} \right) \boldsymbol{I}^\mathcal{S} + 2 \mu^\mathcal{S} \boldsymbol{E}^\mathcal{S}. \label{eq:Prob_14}
\end{equation} 
$\kappa^\mathcal{S}$, $\mu^\mathcal{S}$ and $\boldsymbol{I}^\mathcal{S}$ are the surface dilatational viscosity, the surface shear viscosity and the surface identity tensor, respectively. As for the constitutive law of a Newtonian fluid in the bulk, it can be shown with the help of the second law of thermodynamics that $\kappa^\mathcal{S} \geq 0$ and $\mu^\mathcal{S} \geq 0$ \citep{Gatignol2001}. Furthermore, experimentalists have shown that usually $\kappa^\mathcal{S} > \mu^\mathcal{S}$ in the case of particle-free interfaces \citep[e.g.,][]{Fox1946, Maru1979}. Based on the Boussinesq-Scriven law, we write 
\begin{equation}
\langle\boldsymbol{\sigma}^{\mathcal{S}}\rangle : \boldsymbol{E}^{\mathcal{S}} = (\kappa^{\mathcal{S}} - \mu^{\mathcal{S}}) \left(\langle \boldsymbol{\nabla}^{\mathcal{S}} \cdot \boldsymbol{u}^{\mathcal{S}} \rangle\right)^2 + 2 \mu^{\mathcal{S}} \left(\langle \boldsymbol{E}^\mathcal{S} \rangle\right)^2,\label{eq:Prob_15}
\end{equation}
where we have used the short-hand notation $\left(\langle \boldsymbol{E}^\mathcal{S} \rangle\right)^2 = \langle \boldsymbol{E}^\mathcal{S} \rangle:\langle \boldsymbol{E}^\mathcal{S} \rangle$.

\section{Effective surface viscosities of a particle-laden interface in the dilute limit}\label{sec:Eff}
In this section, we {\color{black} present the strategy used for the calculation of the }effective surface viscosities corresponding to a Boussinesq-Scriven interface with the help {\color{black}of equations (\ref{eq:Prob_12}) and (\ref{eq:Prob_15})}. The solution procedure of equation (\ref{eq:Prob_4}) subject to the boundary conditions from equation (\ref{eq:Prob_7}) is discussed {\color{black} in more detail in appendix \ref{sec:S2a}}. {\color{black} In general, our} solution strategy is based on domain perturbation method. More specifically, we consider the solution of the Stokes equations for a slightly deformed sphere in pure straining flow. To keep the calculation managable, we limit ourselves to the second-order perturbation in a small parameter $\beta$.\\

According to \cite{Brenner1964}, the parametrization of the surface of a slightly deformed sphere can be written as
\begin{equation}
r_\mathcal{P} \approx a + a \beta f(\theta,\varphi) + a \beta^2 g(\theta, \varphi), \label{eq:Eff_1}
\end{equation}
in which $a$ is the radius of the undeformed sphere, $\beta$ a perturbation parameter, and $f(\theta,\varphi)$ and $g(\theta,\varphi)$ two functions of the azimuthal and the polar angle. For a mirror-fused particle consisting of two spherical caps, $\beta = \cos(\alpha)$ is a useful choice \citep{Dorr2015,Dorr2015a, Eigenbrod2018}, and we obtain
\begin{equation}
r_\mathcal{P} \approx a + a \beta \sin(\theta) |\cos(\varphi)| + a \beta^2 ~\frac{\sin^2(\theta) \cos^2(\varphi) - 1}{2}, \label{eq:Eff_2}
\end{equation} 
from which $f(\theta,\varphi) = \sin(\theta) |\cos(\varphi)|$ and $g(\theta, \varphi) = \frac{\sin^2(\theta) \cos^2(\varphi) - 1}{2}$ can be read off. With equation (\ref{eq:Eff_2}) and (\ref{eq:Prob_7}), the boundary condition can be rewritten. Following the procedure first presented by \cite{Brenner1964}, the velocity and pressure field are expanded into a perturbation series in $\beta$ (in which the Stokes equations has to be fulfilled at each order) and each contribution of given order in $\beta$ is Taylor-expanded in $r$ around ($r-a$). After some rearrangement, the following boundary conditions are obtained: 
\begin{eqnarray}
\left. \boldsymbol{u}^{(0)} \right|_a &=& - \boldsymbol{E} \cdot \frac{\boldsymbol{r}}{r} a, \label{eq:Eff_3} \\
\left. \boldsymbol{u}^{(1)} \right|_a &=& - a f(\theta, \varphi) \left( \boldsymbol{E} \cdot \frac{\boldsymbol{r}}{r}  + \left. \frac{\partial \boldsymbol{u}^{(0)}}{\partial r}\right|_a \right), \label{eq:Eff_4}\\
\left. \boldsymbol{u}^{(2)} \right|_a &=& - a f(\theta, \varphi) \left. \frac{\partial \boldsymbol{u}^{(1)}}{\partial r}\right|_a  - a g(\theta, \varphi) \left(\boldsymbol{E} \cdot \frac{\boldsymbol{r}}{r} + \left. \frac{\partial \boldsymbol{u}^{(0)}}{\partial r} \right|_a \right) \nonumber \\
&&  - \frac{a^2}{2} f(\theta,\varphi)^2 \left. \frac{\partial^2 \boldsymbol{u}^{(0)}}{\partial r^2}\right|_a . \label{eq:Eff_5}
\end{eqnarray}
Here, the superscript corresponds to the approximation order in $\beta$. It should be noted that the boundary conditions on the slightly deformed sphere are mapped onto an undeformed sphere. Clearly, the solution of the Stokes equations satisfying boundary condition (\ref{eq:Eff_3}) corresponds to an undeformed sphere in pure straining and is readily known. The boundary condition of the first-order solution is uniquely determined by the zeroth-order solution. To proceed, both functions $f(\theta, \varphi)$ and $g(\theta, \varphi)$ are expanded in surface spherical harmonics \citep{Brenner1964}. For the first-order solution, we therefore expand $f(\theta, \varphi) = \sin(\theta) |\cos(\varphi)| = \sum_{k = 0}^{\infty} f_k$, using a method described in \cite{Byerly1893} or \cite{MacRobert1947}. The terms of this infinite sequence are given through
\begin{eqnarray}
A_{0,k} &=& \frac{2k+1}{4\upi} \int\limits_{0}^{2\upi} \int\limits_{0}^{\upi} f(\theta,\varphi) P_k(\cos(\theta)) \sin(\theta)~\mathrm{d} \theta~\mathrm{d} \varphi, \label{eq:Eff_5b} \\
A_{n,k} &=& \frac{2k+1}{2\upi} \frac{(k-n)!}{(k+n)!} \int\limits_{0}^{2\upi} \int\limits_{0}^{\upi} f(\theta,\varphi) \cos(n \varphi) P_k^n(\cos(\theta)) \sin(\theta)~\mathrm{d} \theta~\mathrm{d} \varphi, \label{eq:Eff_5c} \\
B_{n,k} &=& \frac{2k+1}{2\upi} \frac{(k-n)!}{(k+n)!} \int\limits_{0}^{2\upi} \int\limits_{0}^{\upi} f(\theta,\varphi) \sin(n \varphi) P_k^n(\cos(\theta)) \sin(\theta)~\mathrm{d} \theta~\mathrm{d} \varphi, \label{eq:Eff_5d} \\
f_k &=& A_{0,k} P_k(\cos(\theta)) + \sum\limits_{n=1}^{k} \left(A_{n,k} \cos(n \varphi) + B_{n,k} \sin(n \varphi) \right) P_k^n(\cos(\theta)), \label{eq:Eff_5e}
\end{eqnarray}
in which $P_k(x)$ and $P_k^n(x)$ denote the Legendre polynomials of the first kind and the associated Legendre polynomials of the first kind, respectively. We obtain an infinite sum with $f_{2m -1}(\theta, \varphi) = 0, \forall m = 1,2, \dots$ and $f_{2m}(\theta, \varphi) \neq 0, \forall m = 0,1, \dots$. Even though the first-order stresslet can be calculated exactly, as shown in appendix \ref{sec:S2} for an arbitrary slightly deformed sphere, the boundary condition for the second-order solution has to be approximated due to the presence of the term proportional to the normal gradient of $\boldsymbol{u}^{(1)}$ in equation (\ref{eq:Eff_5}). We discuss this problem in more detail in {\color{black} appendix \ref{sec:Eff_2}}.\\

For convenience, we may also expand the stresslet into a perturbation series:
\begin{equation}
\boldsymbol{S} \approx \boldsymbol{S}^{(0)} + \beta \boldsymbol{S}^{(1)} + \beta^2 \boldsymbol{S}^{(2)}.\label{eq:Eff_6}
\end{equation}
With the help of equation {\color{black}(\ref{eq:Prob_12})} we obtain up to second order in $\beta = \cos(\alpha)$
\begin{equation}
{\color{black} \langle \boldsymbol{\sigma}^\mathcal{S} \rangle} : \boldsymbol{E}^{\mathcal{S}} = \frac{\phi}{2 \pi a^2} \left[\boldsymbol{S}^{(0)} + \beta \boldsymbol{S}^{(1)} + \beta^2 \left(\boldsymbol{S}^{(0)} + \boldsymbol{S}^{(2)} \right) \right] : \boldsymbol{E}, \label{eq:Eff_7}
\end{equation}
if the surface concentration is defined as in equation (\ref{eq:Prob_2}). In contrast, by employing the surface concentration defined by \cite{Lishchuk2009}, we find up to the second order
\begin{equation}
{\color{black} \langle \boldsymbol{\sigma}^\mathcal{S} \rangle} : \boldsymbol{E}^{\mathcal{S}} = \frac{\phi_L}{2 \pi a^2} \left(\boldsymbol{S}^{(0)} + \beta \boldsymbol{S}^{(1)} + \beta^2  \boldsymbol{S}^{(2)} \right) : \boldsymbol{E}. \label{eq:Eff_8}
\end{equation}
Again, equation (\ref{eq:Eff_8}) is used for verification purposes only, since we prefer using $\phi$ for the surface concentration. The double contraction product of the stresslet and the rate of strain tensor is proportional to the square of the surface divergence of the surface velocity as well as to the double contraction of the surface rate of strain tensor. We may write
\begin{eqnarray}
\boldsymbol{S}^{(0)} : \boldsymbol{E} &=& \upi a^3 \mu \left[ C_1^{(0)} \left(\langle \boldsymbol{\nabla}^{\mathcal{S}} \cdot \boldsymbol{u}^{\mathcal{S}} \rangle\right)^2 + C_2^{(0)}  \left(\langle \boldsymbol{E}^\mathcal{S} \rangle\right)^2 \right], \label{eq:Eff_9} \\
\boldsymbol{S}^{(1)} : \boldsymbol{E} &=& \upi a^3 \mu \left[ C_1^{(1)} \left(\langle \boldsymbol{\nabla}^{\mathcal{S}} \cdot \boldsymbol{u}^{\mathcal{S}} \rangle\right)^2 + C_2^{(1)}  \left(\langle \boldsymbol{E}^\mathcal{S} \rangle\right)^2 \right], \label{eq:Eff_10}\\
\left(\boldsymbol{S}^{(0)} + \boldsymbol{S}^{(2)} \right) : \boldsymbol{E} &=& \upi a^3 \mu \left[C_1^{(2)} \left(\langle \boldsymbol{\nabla}^{\mathcal{S}} \cdot \boldsymbol{u}^{\mathcal{S}} \rangle\right)^2 + C_2^{(2)}  \left(\langle \boldsymbol{E}^\mathcal{S} \rangle\right)^2\right], \label{eq:Eff_11}
\end{eqnarray}
from which the constants $C_1^{(i)}$ and $C_2^{(i)}$, with $i = 0,1,2$, {\color{black}{have to be calculated.} We may expand $\mu^\mathcal{S}$ and $\kappa^\mathcal{S}$ as
\begin{eqnarray}
\mu^\mathcal{S} &\approx & \mu^{\mathcal{S}(0)} + \beta \mu^{\mathcal{S}(1)} + \beta^2 \mu^{\mathcal{S}(2)} \label{eq:Eff_12}, \\
\kappa^\mathcal{S} &\approx & \kappa^{\mathcal{S}(0)} + \beta \kappa^{\mathcal{S}(1)} + \beta^2 \kappa^{\mathcal{S}(2)}. \label{eq:Eff_13}
\end{eqnarray}
By comparing coefficients in equations (\ref{eq:Eff_9}) - (\ref{eq:Eff_11}), we find the following relations:
\begin{eqnarray}
\mu^{\mathcal{S}} &=& \frac{\phi \mu a}{4} \left[C_2^{(0)} + \beta C_2^{(1)} + \beta^2 C_2^{(2)}\right], \label{eq:Eff_14} \\
\kappa^{\mathcal{S}} &=& \frac{\phi \mu a}{2} \left[ C_1^{(0)} + \frac{1}{2} C_2^{(0)} + \beta \left( C_1^{(1)} + \frac{1}{2} C_2^{(1)} \right) + \beta^2 \left( C_1^{(2)} + \frac{1}{2} C_2^{(2)}  \right) \right]. \label{eq:Eff_15}
\end{eqnarray}
{\color{black}The details of the computation of the constants $C_1^{(i)}$ and $C_2^{(i)}$, with $i = 0,1,2$, are given in appendix \ref{sec:AppEff}}. The results are summarized in table \ref{tab:t1}.}
\begin{table}
  \begin{center}
\def~{\hphantom{0}}
  \begin{tabular}{lcc}
      $n$  & $C_1^{(n)}$   &  $C_2^{(n)}$ \\[3pt]
       0   & $20/3$ & $20/3$ \\
       1   & $175/12$ & $25/3$ \\
       2   & $\frac{7975319135631907}{496429499940864}$& $\frac{1474603377122345}{248214749970432}$\\
  \end{tabular}
  \caption{Coefficients related to equations (\ref{eq:Eff_9}) - (\ref{eq:Eff_11})}
  \label{tab:t1}
  \end{center}
\end{table}

\section{Discussion \& Application}\label{sec:Diss}
The aim of this section is to discuss the results obtained for the effective surface viscosities. We furthermore compare our theoretical result for the effective surface dilatational viscosity to the {\color{black}semi-analytical} solution obtained in a previous work \citep{Lishchuk2016}. After that, we briefly discuss the application of our theory in the context of numerical simulations. Finally, we utilize our results to study the influence of interfacial particles on the stability of a liquid jet. 

\subsection{Discussion of our results}
With the help of equations (\ref{eq:Eff_12}), (\ref{eq:Eff_13}) as well as {\color{black}the results given in table \ref{tab:t1}, we obtain} for the effective surface viscosities
\begin{eqnarray}
\mu^{\mathcal{S}} &\approx & \frac{5}{3} \mu a \phi \left(1 + \frac{5}{4} \cos(\alpha) + 0.8912 \cos^2(\alpha) \right), \label{eq:Diss_1}\\
\kappa^{\mathcal{S}} &\approx & 5 \mu a \phi \left(1 + \frac{15}{8} \cos(\alpha) + 1.9036 \cos^2(\alpha) \right), \label{eq:Diss_2}
\end{eqnarray}
or when using the surface concentration as defined in \cite{Lishchuk2009}, i.e., equation (\ref{eq:Prob_1}),
\begin{eqnarray}
\mu^{\mathcal{S}} &\approx & \frac{5}{3} \mu a \phi_L \left(1 + \frac{5}{4} \cos(\alpha) -0.1089 \cos^2(\alpha) \right), \label{eq:Diss_3}\\
\kappa^{\mathcal{S}} &\approx & 5 \mu a \phi_L \left(1 + \frac{15}{8} \cos(\alpha) + 0.9036 \cos^2(\alpha) \right). \label{eq:Diss_4}
\end{eqnarray}
\begin{figure}
  \centerline{\includegraphics[width=0.95\textwidth]{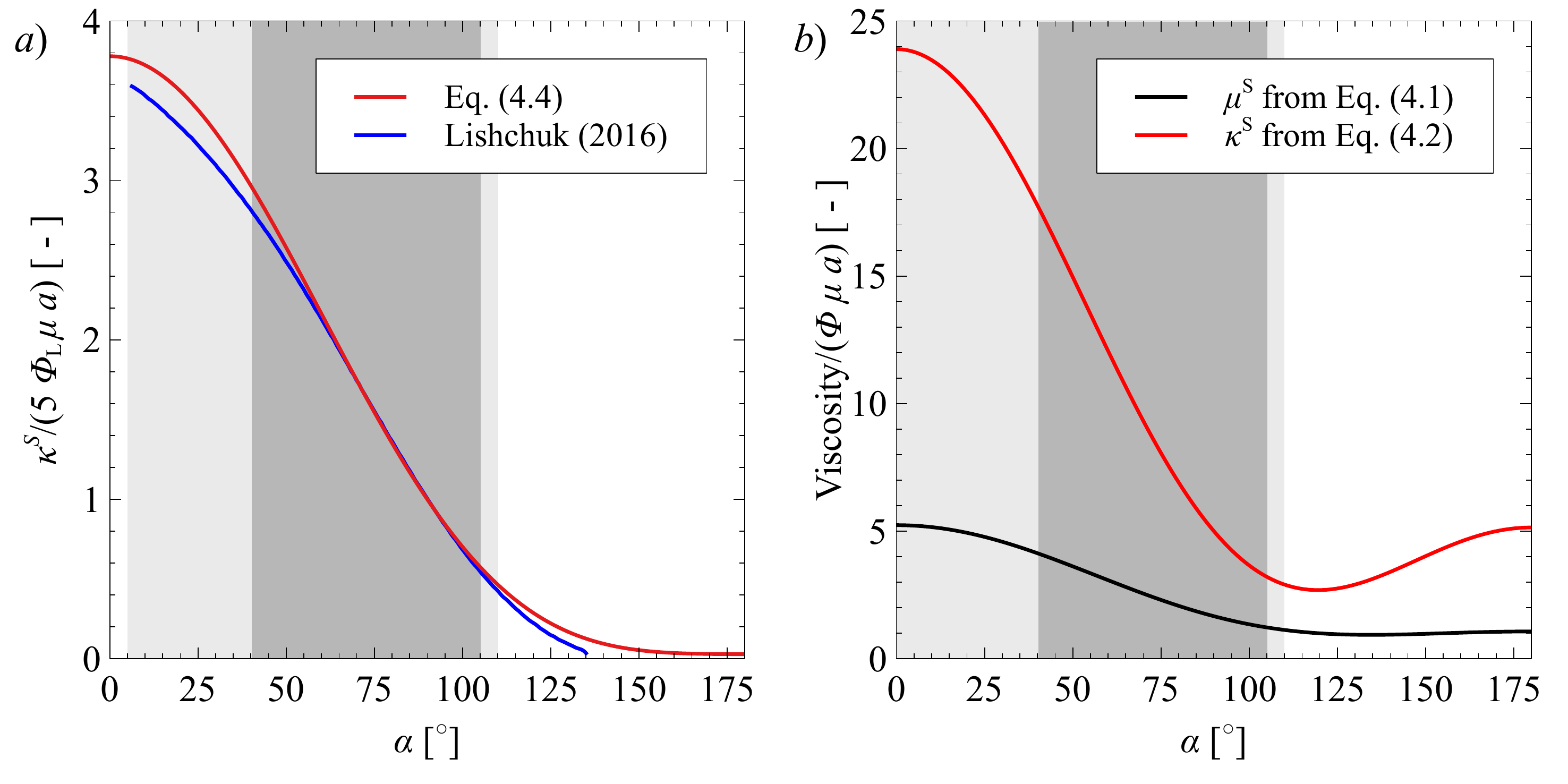}}
  \caption{$a)$ Comparison between the prediction of equation (\ref{eq:Diss_4}) and the {\color{black}semi-analytical} solution obtained in \cite{Lishchuk2016}. The scaled dimensionless dilatational viscosity is shown as a function of the contact angle $\alpha$, where the definition of surface concentration from equation (\ref{eq:Prob_1}) is used. In the dark grey region the deviation between both solutions is smaller than 5 \%, whereas the light grey region corresponds to deviations smaller than 10 \%. $b)$ Dimensionless effective dilatational and shear viscosity as a function of the contact angle. In contrast to part $a)$, the physically correct surface concentration as defined in equation (\ref{eq:Prob_2}) is used. For the grey regions the same convention as in part a) applies.}
\label{fig:f2}
\end{figure}
In figure \ref{fig:f2} $a$), a comparison between equation (\ref{eq:Diss_4}) and the {\color{black}semi-analytical} solution from \cite{Lishchuk2016} is shown. For a broad range of contact angles $\alpha \in [40^\circ; 105^\circ]$ the deviation between both data sets is less than $5$ \% (dark grey region in figure \ref{fig:f2} $a$). The lighter grey areas correspond to a deviation of less than $10$ \%. Similar as in previous studies \citep{Dorr2015, Dorr2015a, Eigenbrod2018}, the domain perturbation method leads to a better agreement with Lishchuk's results for hydrophilic particles than for hydrophobic ones. \\

In figure \ref{fig:f2} $b$), we plot the results of equations (\ref{eq:Diss_1}) and (\ref{eq:Diss_2}) over the contact angle $\alpha$ of the particles. Again, the shaded areas refer to deviations of $5$ \% and $10$ \% between the analytical and the {\color{black}semi-analytical} results for the dilatational viscosity. The dilatational viscosity shows a stronger dependence on the contact angle than the shear viscosity and also has a larger magnitude for all contact angles. Clearly, outside the shaded area towards superhydrophobic contact angles, both viscosities behave differently than expected ($\kappa^\mathcal{S}, \mu^\mathcal{S} \neq 0$, for $\alpha \to 180^\circ$).\\

Our theoretical model is limited to dilute suspensions in which $\phi \ll 1$. By employing Lattice-Boltzmann simulations, \cite{Lishchuk2009} showed that the effective dilatational viscosity for interfacial particles having a contact angle of $\alpha = 90^\circ$ is valid up to surface concentrations of $\phi_L \approx 0.15$. We view this as a rough guideline for validity range of our theory for all contact angles. Compared to volumetric suspensions, in which the classic expression due to Einstein is typically valid up to volume concentrations of 0.05 \citep{Guazzelli2011, Mewis2012}, higher-order corrections in the particle concentration appear less important. However, for fluid interfaces densely packed with particles the interfacial viscosities should show the same divergent behavior as known for volumetric suspensions. For the dilatational viscosity of a particle-laden interface where the particles have a contact angle of $\alpha = 90^\circ$ this was studied by {\color{black}\cite{Lishchuk2014}}. A divergent behavior similar to that of volumetric suspensions was obtained. We therefore hypothesize that our theoretical results for the dilute limit provide a lower bound to the effective viscosities and significantly underestimate these, especially for higher surface concentrations. In other words, our results always represent the minimum influence of interfacial particles, a statement of relevance in different applications, to be further discussed in section \ref{sec:Stability}.\\

Apart from the dimensionless surface viscosities discussed in figure \ref{fig:f2} $b$), two additional dimensionless groups were shown to have an important influence on the evolution of a fluid interface, i.e., the dilatational and shear Boussinesq numbers \citep{Brenner1991}
\begin{eqnarray}
\mathcal{B}_{\mu^\mathcal{S}} &{\color{black}=}& \frac{\mu^\mathcal{S}}{\mu L} {\color{black}= \phi \left(\frac{5}{3} + \frac{25}{12}\cos(\alpha) + 1.4853 \cos^2(\alpha) \right) \frac{a}{L}}, \label{eq:Diss_5} \\
\mathcal{B}_{\kappa^\mathcal{S}} &{\color{black}=}& \frac{\kappa^\mathcal{S}}{\mu L} {\color{black} \phi \left(5 + \frac{75}{8}\cos(\alpha) + 9.5178 \cos^2(\alpha) \right) \frac{a}{L}}. \label{eq:Diss_6} 
\end{eqnarray}
in which $L$ is a characteristic dimension of the flow domain. Further, it might be required to evaluate the ratio of both Boussinesq numbers. Up to $O(\cos^2(\alpha))$, we find
\begin{eqnarray}
\frac{\mathcal{B}_{\kappa^\mathcal{S}}}{\mathcal{B}_{\mu^\mathcal{S}}} = \frac{\kappa^\mathcal{S}}{\mu^\mathcal{S}} = 3 + \frac{15}{8} \cos(\alpha)+0.6936 \cos^2(\alpha). \label{eq:Diss_9}
\end{eqnarray} 
Equation (\ref{eq:Diss_9}) reveals that the ratio of the surface viscosities varies between 5 and 2.5 in the dark shaded region of figure \ref{fig:f2}, reflecting the range of validity of our theory. 

\subsection{Towards numerical simulations of particle-laden interfaces}\label{sec:NumSim}
To make full use of our theoretical results in numerical simulations involving particle-laden interfaces, the isotropic part of the interfacial stress tensor [i.e., the effective interfacial tension \citep{Brenner1991}] has to be considered as well. Following the general idea described by \cite{Bormashenko2013}, in which the effective interfacial tension is defined as the ratio of the total interfacial free energy and the total surface area, we obtain up to first order in the surface concentration $\phi$ and up to second order in $\cos(\alpha)$ (see appendix \ref{sec:S3} for more details):
\begin{eqnarray}
\gamma^\mathcal{S} &=& \gamma_{12} \left[1 - \phi \left(1 + 2 \cos^2(\alpha) \right) \right] + 2 \left(\gamma_{\mathcal{P}(1)} + \gamma_{\mathcal{P}(2)} \right) (1 + \cos^2(\alpha)) \phi, \label{eq:Diss_16}
\end{eqnarray}
in which $\gamma_{12}$ is the interfacial tension between both fluid phases and $\gamma_{\mathcal{P}(\mathrm{i})}$  the interfacial tension between a particle and phase $i$ ($i = 1,2$). The total surface stress tensor $\boldsymbol{\Pi}^\mathcal{S}$ reflecting the presence of a particle-laden fluid interface can therefore be written as
\begin{eqnarray}
\boldsymbol{\Pi}^\mathcal{S} = \gamma^\mathcal{S} \boldsymbol{I}^\mathcal{S} + (\kappa^{\mathcal{S}} - \mu^{\mathcal{S}} ) \left(\boldsymbol{\nabla}^{\mathcal{S}} \cdot \boldsymbol{u}^{\mathcal{S}}\right) \boldsymbol{I}^{\mathcal{S}} + 2 \mu^{\mathcal{S}} \boldsymbol{E}^{\mathcal{S}}. \label{eq:Diss_17}
\end{eqnarray}
The coefficients $\gamma^\mathcal{S}$, $\mu^\mathcal{S}$ and $\kappa^\mathcal{S}$ can be taken from equation (\ref{eq:Diss_16}), (\ref{eq:Diss_1}) and (\ref{eq:Diss_2}), respectively. \\

In a numerical simulation, the transport equations for the fluid interface have to be solved in addition to the transport equations in the bulk. A derivation and discussion of the transport equations in interfacial flows can be found in \cite{Wang2011}, for example. In cases where the evolution of the concentration field of interfacial particles needs to be taken into account, an additional transport equation has be solved, including all relevant forces acting on the particles, such as the drag force \citep{Dorr2015a} or the force due to the curvature of the interface \citep{Leandri2013,Galatola2014}.

\subsection{Stability of a free liquid thread with a particle-laden surface}\label{sec:Stability}
\begin{figure}
  \centerline{\includegraphics[width=0.5\textwidth]{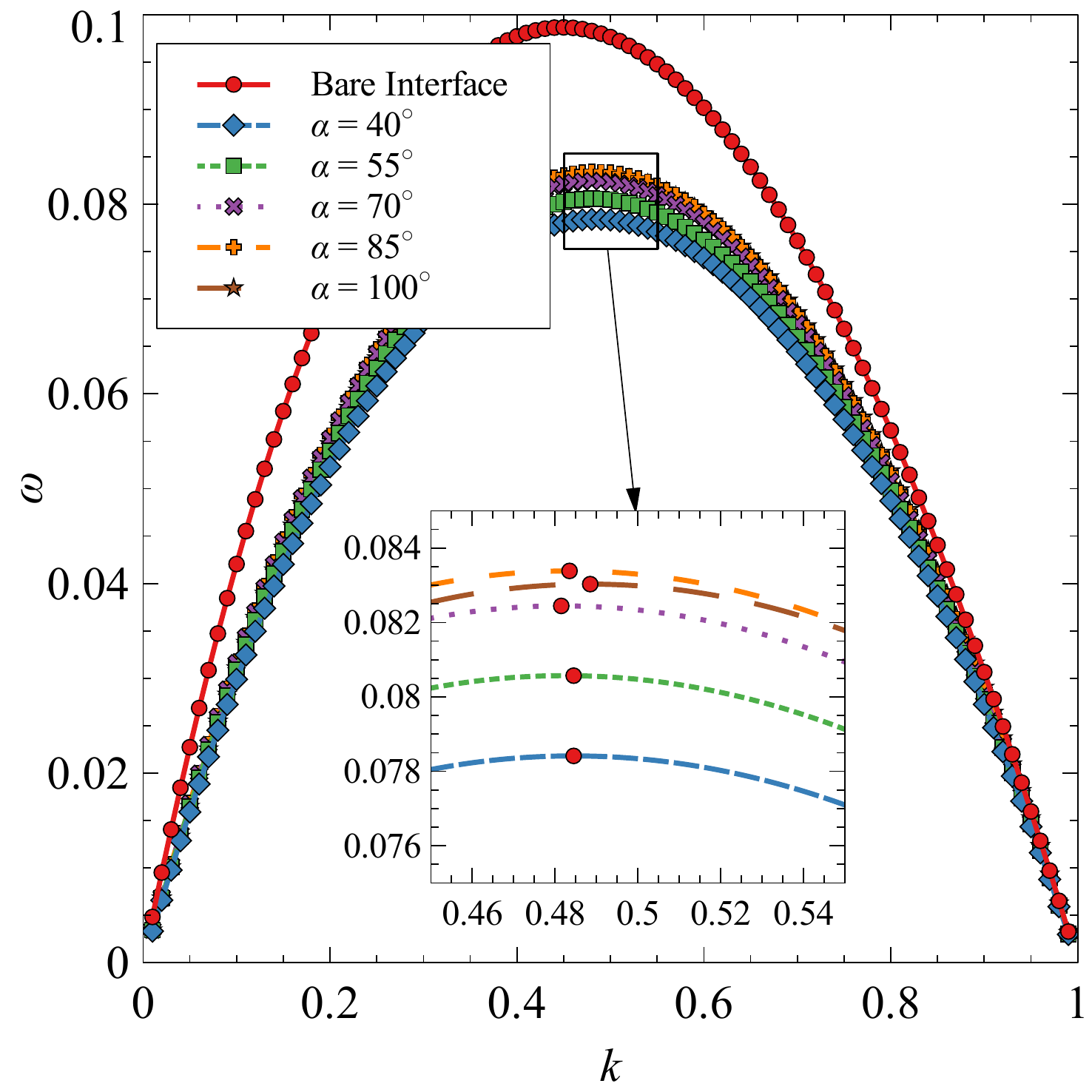}}
  \caption{Plot of growth rate $\omega$ over the wave number $k$ of a particle-laden liquid jet for different contact angles $\alpha$, obtained from the dispersion relation from equation (\ref{eq:Diss_10}) \citep{Martinez-Calvo2018}. For comparison, a particle-free bare interface is added to the plot.}
\label{fig:f7}
\end{figure}

In this subsection, we discuss a potential application of our theoretical analysis in terms of the Rayleigh-Plateau instability of a liquid cylinder having particles attached to its surface. The basis of this is the analysis by \cite{Martinez-Calvo2018}, who studied the temporal axisymmetric instability of a free liquid cylinder coated with insoluble surfactant, including surface elasticity as well as Marangoni effects. We may simplify their dispersion relation [equation (3.26) in \cite{Martinez-Calvo2018}], by neglecting the parameter $\beta$ accounting for the Gibbs elasticity. In addition to our assumptions listed in section \ref{sec:Prob}, it is assumed that the curvature-driven motion of the particles attached to the fluid interface \citep[see][]{Leandri2013,Galatola2014} is much slower than the decay of the liquid cylinder. Correspondingly, we can assume a homogeneous particle distribution at the interface. This is in agreement with the underlying linear stability analysis, since small deviations from a perfectly cylindrical surface are considered. By employing a slightly different notation and neglecting the surface elasticity, we rewrite the dispersion relation into \citep{Martinez-Calvo2018}
\begin{eqnarray}
&&\frac{\mathrm{Re}}{\mathrm{Ca}} \omega^2 F(k) - k^2 (1 - k^2) + k^4 \frac{\mathrm{Ca}}{\mathrm{Re}} \left[ 4 + 6 \mathcal{B}_{\mu^\mathcal{S}}+ \frac{1 - k^2}{\omega} \left(\mathcal{B}_{\mu^\mathcal{S}} + \mathcal{B}_{\kappa^\mathcal{S}} \right)  - 2 \mathcal{B}_{\kappa^\mathcal{S}} (1 + 2 \mathcal{B}_{\mu^\mathcal{S}}) \right]  \nonumber \\ 
&&   (F(k)-F(\tilde{k})) + \omega k^2 \left[2 (\mathcal{B}_{\mu^\mathcal{S}} - \mathcal{B}_{\kappa^\mathcal{S}}) F(k) + (\mathcal{B}_{\mu^\mathcal{S}} + \mathcal{B}_{\kappa^\mathcal{S}}) (F(k) F(\tilde{k})+1) \right. \nonumber \\
&& \Big. + 2 (F(k) -1) \Big]  = 0, \label{eq:Diss_10}
\end{eqnarray} 
with
\begin{eqnarray}
\tilde{k} = \sqrt{k^2 + \frac{\mathrm{Re}}{\mathrm{Ca}} \omega}\quad \mathrm{and} \quad F(x) = x \frac{I_0(x)}{I_1(x)}, \label{eq:Diss_11}
\end{eqnarray}
in which $\mathrm{Re} = \rho U R/\mu$, $\mathrm{Ca} = \mu U/\gamma^\mathcal{S}$, $k$, $\omega$, $R$ and $I_n(x)$ are the Reynolds number, the capillary number, the wavenumber, the growth-rate, the radius of the unperturbed liquid jet and the $n$-th order modified Bessel function of the first kind \citep{Martinez-Calvo2018}, respectively. It should be noted that the capillary number depends on the contact angle through the effective surface tension $\gamma^\mathcal{S}$, presented in the previous subsection. Consequently, changing the wettability of the interfacial particles under constant flow conditions changes the ratio of Reynolds and capillary number. We make use of the Boussinesq numbers from equation (\ref{eq:Diss_5}) and (\ref{eq:Diss_6}) with $L = R$, which reveals that the influences of the effective surface viscosities can only be considered asymptotically, since $\phi$ and $a/R$ are both small parameters. Note that the latter ratio needs to be a small to satisfy assumption (\ref{itm:3}). In what follows we set $\phi = 0.15$ and $a/R = 0.1$. The combination of a highly viscous liquid and sufficiently small characteristic velocities $U$ should result in $\mathrm{Re} \ll 1$ and $\mathrm{Ca} \ll 1$, in line with the assumptions made to compute the effective viscosities. We therefore consider a silicone oil thread (100 cSt, $\rho = 0.96$ g/ml, $\gamma_{12} = 20$ mN/m) with a radius $R = 1$ mm. Based on that, for $U = 10$ mm/s we find $\mathrm{Re} = 0.1$. Assuming a small particle concentration, the effective surface tension is comparable to the surface tension of the bare silicone oil. The capillary number is then $\mathrm{Ca} = O(10^{-2})$. Equation (\ref{eq:Diss_16}) requires knowledge about the interfacial tensions between the particles and the surrounding media. We consider polystyrene particles for which $\gamma_{\mathcal{P}(\mathrm{air})} = 35$ mN/m \citep{Shimizu2000}. When we vary the contact angle, we keep $\gamma_{\mathcal{P}(\mathrm{air})}$ fixed to that value. The interfacial tension between the particle and silicone oil is then calculated using Young's law, see appendix \ref{sec:S3}. After inserting the Boussinesq numbers and the capillary number into equation (\ref{eq:Diss_10}) and (\ref{eq:Diss_11}), we employ a Taylor series up to the second order in $\cos(\alpha)$ and up to the first order in $\phi$, to be consistent with our previous derivation. Subsequently, we solve equation (\ref{eq:Diss_10}) numerically, using Newton's method. The results are shown in figure \ref{fig:f7}. \\

The influence of surface particles on the dispersion relation is such that the growth rate of perturbations reduces when particles are present. This qualitative observation agrees with various experiments performed in different contexts \citep[e.g.,][]{Binks2002,Aveyard2003,Dickinson2010,Wu2016,DeCorato2018}. A surface concentration of $\phi = 0.15$ yields a growth rate reduced by about 20\% compared to a particle-free surface. In this particular example, interfacial particles with a contact angle deviating from $\alpha = 90^\circ$ slow down the growth of modes more significantly than particles with a contact angle close to $90^\circ$. Owing to the presence of particles, the critical wave number increases by about 7\% for all contact angles considered. It could be tempting to use the computed surface viscosity values for cases with higher surface concentrations. As explained above, we expect that our theory underestimates the influence of interfacial particles for larger values of $\phi$. In other words, in such cases the presence on interfacial particles is expected to influence the dispersion relation even more significantly than visible in figure \ref{fig:f7}.

\section{Conclusion}\label{sec:Conc}
In summary, we have studied the dissipation effects due to spherical particles adsorbed at the interface between two fluids with large viscosity contrast and computed the effective interfacial dilatational and shear viscosity. We limited our study to flat interfaces and small Reynolds numbers. Furthermore, the interfacial suspension was assumed to be dilute, which allows neglecting the hydrodynamic, capillary and Brownian interactions between particles. {\color{black} Throughout this study, we made use of symmetry arguments stating that the physical situation of a particle attached to a flat interface is equivalent to that of a mirror-fused particle in the bulk of the fluid with higher viscosity. We applied the stress-averaging method by \cite{Batchelor1970} to interfacial suspensions and found that the effective surface viscosities are solely determined by the stresslet acting on the mirror-fused particle.}\\

The shape of the mirror-fused particle was modeled based on a domain perturbation method, where the contact angle $\alpha$ of the more viscous fluid on the particle surface defines the particle shape. By applying the Lorentz reciprocal theorem \citep{Lorentz1896a} and solving the Stokes equations for the mirror-fused particle in a pure straining flow up to first order in the perturbation parameter $\beta = \cos(\alpha)$, we computed both interfacial viscosities of a Boussinesq-Scriven interface up to second order in $\beta$. We found that the dilatational viscosity is more sensitive to the contact angle than the shear viscosity. Further, the magnitude of the dilatational viscosity was shown to be larger than the magnitude of the shear viscosity, which is in agreement with experimental results obtained for particle-free fluid interfaces \citep{Fox1946,Maru1979}. Usually, in the corresponding articles clean interfaces are considered, but recent results indicate that in practice, surface contaminations are very difficult to avoid \citep{Landel2017}. By comparing our results with the {\color{black}semi-analytical} results of \cite{Lishchuk2016}, we found that our perturbation approach agrees well with Lishchuk's data (deviation less than 5 \%) in a contact angle range between $40^\circ$ and $105^\circ$. To provide a complete picture of the interfacial stresses, we presented an expression for the effective interfacial tension as a function of the contact angle. \\

An application of our theory was discussed in terms of the decay of a liquid cylinder whose surface is decorated with particles. Based on the theory by \cite{Martinez-Calvo2018} and considering a particle surface concentration of 15\%, we found that interfacial particles slow down the decay by about 20\% compared to a bare interface. \\ 

The results may prove useful in different contexts. First, they could help understanding low-concentration particle-laden interfacial flows, especially the role of the contact angle in these flows. Second, in a similar way as with theories for the effective viscosity of a bulk suspension, our findings could pave the way to follow-up studies considering higher particle concentrations. This could enable understanding the hydrodynamics of systems as complex as Pickering emulsions and liquid marbles.


\appendix
\section{Contributions of $\boldsymbol{U}^{\mathrm{eff}}$ and $\boldsymbol{\omega}^{\mathrm{eff}}$ in the evaluation of the effective surface viscosities}\label{sec:S1}

We start our analysis with equation (\ref{eq:Prob_5}) and derive the corresponding boundary conditions for a slightly deformed sphere up to second order in the small parameter $\beta$, according to the solution procedure described in section \ref{sec:Eff}. We therefore consider a surface parametrization as given in equation (\ref{eq:Eff_1}). The boundary conditions are then given by (more details in appendix \ref{sec:S2})
\begin{eqnarray}
\left. \boldsymbol{u}^{(0)}\right|_{r = a} &=& - \boldsymbol{E} \cdot \frac{\boldsymbol{r}}{r} a, \label{eq:SA_1} \\
\left. \boldsymbol{u}^{(1)}\right|_{r = a} &=& \left( \boldsymbol{U}^{\mathrm{eff},(1)} + \boldsymbol{\omega}^{\mathrm{eff},(1)} \times \frac{\boldsymbol{r}}{r} a \right)  - a f(\theta, \varphi) \left(\boldsymbol{E} \cdot \frac{\boldsymbol{r}}{r} + \left. \frac{\partial \boldsymbol{u}^{(0)}}{\partial r}\right|_{r = a} \right), \label{eq:SA_2} \\
\left. \boldsymbol{u}^{(2)}\right|_{r = a} &=& \left( \boldsymbol{U}^{\mathrm{eff},(2)} + \boldsymbol{\omega}^{\mathrm{eff},(2)} \times \frac{\boldsymbol{r}}{r} a \right) - a f(\theta, \varphi) \left. \frac{\partial \boldsymbol{u}^{(1),\mathrm{T}}}{\partial r}\right|_{r = a} \nonumber \\
&+& a f(\theta,\varphi) \left(\boldsymbol{\omega}^{\mathrm{eff},(1)} \times \frac{\boldsymbol{r}}{r} - \left. \frac{\partial \boldsymbol{u}^{(1),\mathrm{R}}}{\partial r}\right|_{r = a} \right) - a f(\theta, \varphi) \left. \frac{\partial \boldsymbol{u}^{(1),\mathrm{S}}}{\partial r}\right|_{r = a} \nonumber \\
&-& a g(\theta, \varphi) \left(\boldsymbol{E} \cdot \frac{\boldsymbol{r}}{r} + \left. \frac{\partial \boldsymbol{u}^{(0)}}{\partial r}\right|_{r = a} \right) - \frac{a^2}{2} f(\theta,\varphi)^2 \left. \frac{\partial^2 \boldsymbol{u}^{(0)}}{\partial r^2}\right|_{r = a}, \label{eq:SA_3}
\end{eqnarray}
in which we used the expansions $\boldsymbol{U}^{\mathrm{eff}} = \beta \boldsymbol{U}^{\mathrm{eff},(1)} + \beta^2 \boldsymbol{U}^{\mathrm{eff},(2)}$, $\boldsymbol{\omega}^{\mathrm{eff}} = \beta \boldsymbol{\omega}^{\mathrm{eff},(1)} + \beta^2 \boldsymbol{\omega}^{\mathrm{eff},(2)}$, as well as $\boldsymbol{u}^{(1)} = \boldsymbol{u}^{(1),\mathrm{T}} + \boldsymbol{u}^{(1),\mathrm{R}} + \boldsymbol{u}^{(1),\mathrm{S}}$, i.e. the decomposition of the first order velocity field in translational, rotational and pure straining contributions. As shown in section \ref{sec:Prob_1}, the stresslet takes a major role in the calculation of the effective interfacial viscosities. We will therefore concentrate on the influence of the boundary conditions on the stresslet.\\

The boundary condition for the first-order velocity field is a superposition of the boundary conditions referring to an undeformed sphere translating and rotating with $\boldsymbol{U}^{\mathrm{eff},(1)}$ and $\boldsymbol{\omega}^{\mathrm{eff},(1)}$, respectively, and a slightly deformed sphere in a pure straining flow. However, since the translational and rotational motion of an undeformed sphere do not contribute to the stresslet, it is sufficient to consider the pure-straining boundary condition for a slightly deformed sphere in the first-order contribution. Analogously, for the second order in $\beta$, i.e., equation (\ref{eq:SA_3}), the first term in round brackets on the right-hand side does not need to be considered, since this also refers to an undeformed sphere in a constant translational and rotational velocity field. The following two terms correspond to a slightly deformed sphere up to first order of $\beta$ in pure translational ($\boldsymbol{U}^{\mathrm{eff},(1)}$) and rotational ($\boldsymbol{\omega}^{\mathrm{eff},(1)}$) motion  \citep[for comparison, see equations (3.9) \& (4.3) in][]{Brenner1964}, that contributes to the stresslet whenever at least one of these velocities is nonvanishing. The remaining terms correspond to a second-order slightly deformed sphere in a pure straining flow. In the following, we calculate the first-order effective velocities on the basis of a force and torque balance for the most general parametrization of a slightly deformed sphere. \\

To begin with, we expand the force and torque in a power series in $\beta$ up to first order, leading to
\begin{eqnarray}
\boldsymbol{F} &\approx & \boldsymbol{F}^{(0)} + \beta \boldsymbol{F}^{(1)}, \label{eq:SA_4} \\
\boldsymbol{T} &\approx & \boldsymbol{T}^{(0)} + \beta \boldsymbol{T}^{(1)}. \label{eq:SA_5}
\end{eqnarray}
For an undeformed sphere it is readily known that the rotation (translation) does not contribute to the force (torque) acting on it. We have
\begin{eqnarray}
\boldsymbol{F}^{(0)} = -6 \upi \mu a \boldsymbol{U}^{\mathrm{eff}}, \label{eq:SA_6} \\
\boldsymbol{T}^{(0)} = -8 \upi \mu a^3 \boldsymbol{\omega}^{\mathrm{eff}}. \label{eq:SA_7}
\end{eqnarray}
In the first order of $\beta$, the force and torque can be decomposed in a translational, rotational and straining contribution. Using the solution for a translating and rotating slightly deformed sphere obtained by \citet{Brenner1964} and the solution procedure discussed in appendix \ref{sec:S2} for a sphere in pure straining flow, the force and torque can be calculated using the following relations \citep{Brenner1964,Kim2013}
\begin{eqnarray}
\boldsymbol{F} &=& - 4 \upi \boldsymbol\nabla(r^3 p_{-2}), \label{eq:SA_8} \\ 
\boldsymbol{T} &=& - 8 \upi \mu \boldsymbol\nabla(r^3 \chi_{-2}), \label{eq:SA_9}
\end{eqnarray}
in which $p_{-2}$ and $\chi_{-2}$ are solid spherical harmonics of order $-2$ and part of Lamb's general solution (see appendix \ref{sec:S2} for more details). After some algebra, the following identities are obtained
\begin{eqnarray}
\boldsymbol{F}^{(1)} &=& \boldsymbol{F}^{\mathrm{T}, (1)} + \boldsymbol{F}^{\mathrm{R}, (1)}+ \boldsymbol{F}^{\mathrm{S}, (1)}, \label{eq:SA_10}
\end{eqnarray}
with
\begin{eqnarray}
\boldsymbol{F}^{\mathrm{T}, (1)} &=& -6\upi \mu a \left( \boldsymbol{U}^{\mathrm{eff}} f_0 - \frac{1}{10} \boldsymbol{U}^{\mathrm{eff}} \cdot \boldsymbol{\nabla} \boldsymbol{\nabla} (r^2 f_2) \right), \label{eq:SA_11}\\
\boldsymbol{F}^{\mathrm{R}, (1)} &=& -6 \upi \mu a^2 \boldsymbol{\omega}^{\mathrm{eff}} \times \boldsymbol{\nabla}(r f_1), \label{eq:SA_12}\\
\boldsymbol{F}^{\mathrm{S}, (1)} &=& -4\upi \mu a^2 \left(\frac{1}{14}\boldsymbol{E}:\boldsymbol{\nabla}\boldsymbol{\nabla}\boldsymbol{\nabla}(r^3 f_3) - \frac{3}{2} \boldsymbol{E}\cdot \boldsymbol{\nabla}(r f_1) \right),\label{eq:SA_13}
\end{eqnarray}
and
\begin{eqnarray}
\boldsymbol{T}^{(1)} &=& \boldsymbol{T}^{\mathrm{T}, (1)} + \boldsymbol{T}^{\mathrm{R}, (1)}+ \boldsymbol{T}^{\mathrm{S}, (1)}, \label{eq:SA_14}
\end{eqnarray}
with
\begin{eqnarray}
\boldsymbol{T}^{\mathrm{T}, (1)} &=& 6 \upi a^2 \mu \boldsymbol{U}^{\mathrm{eff}} \times \boldsymbol{\nabla} (r f_1), \label{eq:SA_15}\\
\boldsymbol{T}^{\mathrm{R}, (1)} &=& -24 \upi a^3 \mu \left(\boldsymbol{\omega}^{\mathrm{eff}} f_0 - \frac{1}{10} \boldsymbol{\omega}^{\mathrm{eff}}\cdot \boldsymbol{\nabla}\boldsymbol{\nabla}(r^2 f_2) \right), \label{eq:SA_16}\\
\boldsymbol{T}^{\mathrm{S}, (1)} &=& 4 \upi \mu a^3 \boldsymbol{\nabla} \times \left(\boldsymbol{E} \cdot \boldsymbol{\nabla}(r^2 f_2) \right), \label{eq:SA_17}
\end{eqnarray}
in which $f_k$ [$f_k = f_k(\theta,\varphi)$] is the $k$-th partial sum of the spherical-harmonics expansion of the shape function $f(\theta,\varphi)$ [see equation (\ref{eq:Eff_1})]. After writing $\boldsymbol{U}^{\mathrm{eff}} = \beta \boldsymbol{U}^{\mathrm{eff},(1)}$, $\boldsymbol{\omega}^{\mathrm{eff}} = \beta \boldsymbol{\omega}^{\mathrm{eff},(1)}$ and inserting into equations (\ref{eq:SA_6}), (\ref{eq:SA_7}) \& (\ref{eq:SA_10}) - (\ref{eq:SA_17}) we find that the force and torque balance is satisfied up to $O(\beta)$ if
\begin{eqnarray}
\boldsymbol{U}^{\mathrm{eff},(1)} &=& a \left(\boldsymbol{E} \cdot \boldsymbol{\nabla}(r f_1) - \frac{1}{21} \boldsymbol{E}:\boldsymbol{\nabla}\boldsymbol{\nabla}\boldsymbol{\nabla}(r^3 f_3) \right), \label{eq:SA_18} \\
\boldsymbol{\omega}^{\mathrm{eff},(1)} &=& \frac{1}{2} \boldsymbol{\nabla} \times \left( \boldsymbol{E} \cdot \boldsymbol{\nabla}(r^2 f_2)\right). \label{eq:SA_19}
\end{eqnarray}
It is worth mentioning that these results can be used to determine the trajectory of a slightly deformed sphere in a linear flow field. As already mentioned in section \ref{sec:Eff}, the expansion $f(\theta,\varphi) = \sin(\theta) |\cos(\varphi)| = \sum_{k = 0}^{\infty} f_k$  only receives contributions from terms with $k = 0$ or $k$ even. Consequently, it follows that the first-order order effective translational velocity vanishes. With equation (\ref{eq:Prob_6}) and $f_2 = -5/32 \left[3 \cos^2(\theta)-1 + 3 \left(\cos^2(\theta) -1 \right) \cos(2 \varphi) \right]$ it follows that $\boldsymbol{\omega}^{\mathrm{eff},(1)} = 0$. Consequently, up to second order in $\beta$, the effective translational and rotational velocity does not need to be considered, i.e., it is sufficient to take the pure straining boundary condition into account. \\

Even though the effective translational and rotational velocities are zero in the present case, it might be useful for future studies to take the effect of a non-vanishing effective particle motion on the stresslet into account. Using
\begin{eqnarray}
\boldsymbol{S} &\approx & \boldsymbol{S}^{(0)} + \beta \boldsymbol{S}^{(1)}\label{eq:SA_19b}
\end{eqnarray}
and
\begin{eqnarray}
\boldsymbol{S}^{(1)} = \boldsymbol{S}^{\mathrm{T}, (1)} + \boldsymbol{S}^{\mathrm{R}, (1)} + \boldsymbol{S}^{\mathrm{S}, (1)}, \label{eq:SA_20}
\end{eqnarray}
we find
\begin{eqnarray}
\boldsymbol{S}^{(0)} &=& \frac{20}{3} \upi \mu a^3 \boldsymbol{E}, \label{eq:SA_21} \\
\boldsymbol{S}^{\mathrm{T}, (1)} &=& 2 \upi \mu a^2 \left[\frac{1}{7} \boldsymbol{U}^{\mathrm{eff}} \cdot \boldsymbol{\nabla} \boldsymbol{\nabla} \boldsymbol{\nabla} (r^3 f_3) + \boldsymbol{U}^{\mathrm{eff}} \cdot \boldsymbol{\nabla} (r f_1) \boldsymbol{I} \right. \nonumber \\
&-& \left. \frac{3}{2} \left(\boldsymbol{U}^{\mathrm{eff}} \boldsymbol{\nabla} (r f_1) + \boldsymbol{\nabla} (r f_1) \boldsymbol{U}^{\mathrm{eff}}  \right) \right],\label{eq:SA_22} \\
\boldsymbol{S}^{\mathrm{R}, (1)} &=& - 2\upi \mu a^3 \left[\boldsymbol{\nabla} (\boldsymbol{\omega}^{\mathrm{eff}} \times \boldsymbol{\nabla}(r^2 f_2)) + \boldsymbol{\nabla} (\boldsymbol{\omega}^{\mathrm{eff}} \times \boldsymbol{\nabla}(r^2 f_2))^T  \right]. \label{eq:SA_23}
\end{eqnarray}
The term corresponding to pure straining motion, $\boldsymbol{S}^{\mathrm{S},(1)}$, is derived in appendix \ref{sec:S2}. It should be noted that with the help of the boundary condition up to $O(\beta^2)$ together with equations (\ref{eq:SA_20}) - (\ref{eq:SA_23}) and the reciprocal theorem [equation (\ref{eq:Eff_23})], the stresslet for an arbitrary slightly deformed sphere can be calculated up to second order in $\beta$.

{\color{black}\section{Underlying method for the computation of the effective interfacial viscosities} \label{sec:S2a}
The purpose of this appendix is to introduce the underlying method for the computation of the effective interfacial viscosities, which is closely related to the work by \cite{Batchelor1970}. We refer to figure \ref{fig:f1} (top left \& bottom) and relate the suspension stresses inside the bulk (subscript $\mathcal{B}$) to effective interfacial properties (subscript $\mathcal{S}$). \\

To start with, we compute the volume averaged stresses $\langle \boldsymbol{\Sigma}_i \rangle_V$ referring to the bulk containing mirror-fused particles ($i = \mathcal{B}$, see bottom of figure \ref{fig:f1}) and the configuration reflecting effective interfacial suspension properties ($i = \mathcal{S}$, see top left of figure \ref{fig:f1}). The latter contains a Boussinesq-Scriven interface that is characterized by non-vanishing surface stresses. This leads to
\begin{eqnarray}
\langle \boldsymbol{\Sigma}_\mathcal{B} \rangle_V &=& \frac{1}{V_\mathcal{B}} \int\limits_{V_\mathcal{B}} \boldsymbol{\Pi}~\mathrm{d}V, \label{eq:SBa_1}\\
\langle \boldsymbol{\Sigma}_\mathcal{S} \rangle_V &=& \frac{1}{V_\mathcal{S}} \left[ \int\limits_{V_\mathcal{S}} \boldsymbol{\Pi}~\mathrm{d}V + \int\limits_{A} \boldsymbol{\sigma}^\mathcal{S}~\mathrm{d}S \right], \label{eq:SBa_2}
\end{eqnarray}
in which $V_i$, $A$, $\boldsymbol{\Pi}$ and $\boldsymbol{\sigma}^\mathcal{S}$ are the total volume ($i = \mathcal{B}, \mathcal{S}$), the area of the Boussinesq-Scriven interface, the total bulk stress and the deviatoric interfacial stress, respectively. Considering figure \ref{fig:f1} it follows that $V_\mathcal{B} = 2~V_\mathcal{S}$. \cite{Batchelor1970} has shown that the right-hand side of equation (\ref{eq:SBa_1}) can be rewritten into a superposition of terms due to the applied velocity field and an additional term reflecting the stresses due the presence of particles $\boldsymbol{\Sigma}_\mathcal{B}^\mathcal{P}$:
\begin{equation}
\langle \boldsymbol{\Sigma}_\mathcal{B} \rangle_V = -\langle p \rangle_V \boldsymbol{I} + 2 \mu \langle \boldsymbol{E} \rangle_V + \frac{1}{V_\mathcal{B}}\boldsymbol{\Sigma}_\mathcal{B}^\mathcal{P}, \label{eq:SBa_3}
\end{equation}
where $\boldsymbol{E}$ is the rate of strain tensor (see equation \ref{eq:Prob_4b}) and $p$ the pressure field corresponding to the applied velocity field. The same rationale can be applied to the formulation relying on the Boussinesq-Scriven interface, assuming that the presence of particles solely effects the effective rheological properties of the fluid interface. We have
\begin{equation}
\langle \boldsymbol{\Sigma}_\mathcal{S} \rangle_V = -\langle p \rangle_V \boldsymbol{I} + 2 \mu \langle \boldsymbol{E} \rangle_V + \frac{1}{V_\mathcal{S}} \int\limits_{A} \boldsymbol{\sigma}^\mathcal{S}~\mathrm{d}S. \label{eq:SBa_4}
\end{equation}
To relate the surface stress to corresponding volumetric quantities, we utilize the two different expressions for the volume averaged stress to compute the total energy dissipation. More precisely, the total energy dissipation due to the Boussinesq-Scriven interface coincides with the energy dissipation due to mirror-fused particles suspended in a bulk. Recalling that $\boldsymbol{E}$ and $\boldsymbol{E}^\mathcal{S}$ are constant tensors, the energy dissipation of the Boussinesq-Scriven interface is the double contraction (defined by $:$) of the last term on the right-hand side of (\ref{eq:SBa_4}) with $\boldsymbol{E}^\mathcal{S}$, while the last term on the right-hand side of (\ref{eq:SBa_3}) double contracted with $\boldsymbol{E}$ leads to the volumetric equivalent. We find, after computing the area average of the interfacial energy dissipation
\begin{eqnarray}
\langle \boldsymbol{\sigma}^\mathcal{S}\rangle : \boldsymbol{E}^\mathcal{S} = \frac{1}{2 A} \boldsymbol{\Sigma}_\mathcal{B}^\mathcal{P} : \boldsymbol{E}. \label{eq:SBa_5}
\end{eqnarray}  
Making further use of the results obtained by \cite{Batchelor1970} on the relationship between the particle stresses $\boldsymbol{\Sigma}_\mathcal{B}^\mathcal{P}$ and the stresslet, equation (\ref{eq:Prob_10}) is obtained.}

\section{Solution of the Stokes equations for a slightly deformed sphere [$O(\beta)$] in a pure straining flow}\label{sec:S2}
In this section, we derive the solution of the Stokes equations together with the boundary conditions of equation (\ref{eq:Prob_7}) on the basis of Lamb's general solution \citep[e.g.,][]{Happel2012}. Although in the main text we only use the solution from this appendix in the context of mirror-fused particles, we keep the calculation as general as possible, i.e. we use the parametrization of an arbitrary slightly deformed sphere. According to \cite{Brenner1964}, we may limit our analysis to the first-order solution in the perturbation parameter $\beta$. \\
The general parametrization of the surface of a slightly deformed sphere is given by
\begin{equation}
r_\mathcal{P} = a + \beta a f(\theta,\varphi), \label{eq:SB_1}
\end{equation}
in which $\beta$ is assumed to be a small parameter. Equation (\ref{eq:Prob_7}) then reads
\begin{equation}
\left. \boldsymbol{u}\right|_{r_\mathcal{P}} = - \boldsymbol{E} \cdot \frac{\boldsymbol{r}}{r} a \left[1 + \beta f(\theta,\varphi) \right]. \label{eq:SB_2}
\end{equation}
It is further assumed that the velocity field can be expanded in a perturbation series as
\begin{equation}
\boldsymbol{u} = \sum\limits_{n = 0}^{\infty} \beta^n \boldsymbol{u}^{(n)}. \label{eq:SB_3}
\end{equation}
Correspondingly, every velocity field $\boldsymbol{u}^{(n)}$ has to satisfy the Stokes equations. Every contribution $\boldsymbol{u}^{(n)}$ is now expanded into a Taylor series around $(r - a)$. In the same way as in \cite{Brenner1964}, we obtain the following zeroth- and first-order boundary conditions
\begin{eqnarray}
\left. \boldsymbol{u}^{(0)}\right|_{r = a} &=& - \boldsymbol{E} \cdot \frac{\boldsymbol{r}}{r} a, \label{eq:SB_4} \\
\left. \boldsymbol{u}^{(1)}\right|_{r = a} &=& - a f(\theta, \varphi) \left(\boldsymbol{E} \cdot \frac{\boldsymbol{r}}{r} + \left. \frac{\partial \boldsymbol{u}^{(0)}}{\partial r}\right|_{r = a} \right). \label{eq:SB_5}
\end{eqnarray}
The former boundary condition refers to a sphere in a pure straining flow, for which the solution is readily known \citep[e.g.,][]{Guazzelli2011}:
\begin{eqnarray}
\boldsymbol{u}^{(0)} &=& - \frac{5 a^3}{2} \frac{\boldsymbol{r} \left(\boldsymbol{r} \cdot \boldsymbol{E} \cdot \boldsymbol{r} \right)}{r^5} - \frac{a^5}{2} \left( \frac{\boldsymbol{E} \cdot \boldsymbol{r} +  \boldsymbol{r}\cdot \boldsymbol{E}}{r^5} - \frac{5 \boldsymbol{r} \left(\boldsymbol{r} \cdot \boldsymbol{E} \cdot \boldsymbol{r} \right)}{r^7} \right), \label{eq:SB_6}\\
p^{(0)} &=& -5 \mu a^3 \frac{\boldsymbol{r} \cdot \boldsymbol{E} \cdot \boldsymbol{r}}{r^5}. \label{eq:SB_7}
\end{eqnarray}
Equation (\ref{eq:SB_5}) then leads to
\begin{equation}
\left. \boldsymbol{u}^{(1)}\right|_{r = a} = 5 a f(\theta, \varphi) \left(\frac{\left(\boldsymbol{r} \cdot \boldsymbol{E} \cdot \boldsymbol{r} \right)}{r^2} {\color{black}\boldsymbol{r}} - \boldsymbol{E} \cdot \frac{\boldsymbol{r}}{r} \right). \label{eq:SB_8} 
\end{equation}
According to \cite{Brenner1964}, the following three terms have to be evaluated and written in terms of three different surface spherical harmonics $X_n, Y_n$ and $Z_n$, each of which is defined with respect to a specific angular function:
\begin{eqnarray}
\sum\limits_{n = 1}^{\infty} X_n &=& \frac{\boldsymbol{r}}{r} \cdot \left. \boldsymbol{u}^{(1)}\right|_a = 0, \label{eq:SB_9} \\
\sum\limits_{n = 1}^{\infty} Y_n &=&  - r \boldsymbol{\nabla} \cdot \left( \left. \boldsymbol{u}^{(1)}\right|_a \right) = 5 a \left(\boldsymbol{\nabla} \left(f(\theta, \varphi)\right) \cdot \boldsymbol{E} \cdot \boldsymbol{r} - 3 f(\theta,\varphi) \frac{\boldsymbol{r} \cdot \boldsymbol{E} \cdot \boldsymbol{r}}{r^2} \right), \label{eq:SB_10} \\
\sum\limits_{n = 1}^{\infty} Z_n &=& \boldsymbol{r} \cdot \boldsymbol{\nabla} \times \left( \left. \boldsymbol{u}^{(1)}\right|_a \right) =5 a \left(\boldsymbol{E} \cdot \frac{\boldsymbol{r}}{r}\right)\cdot \boldsymbol{\nabla} \times \left(f(\theta,\varphi) \boldsymbol{r}\right) \label{eq:SB_11}.
\end{eqnarray}
It follows directly that $X_n = 0$, $\forall n$. We now expand $f(\theta, \varphi) = \sum_{k = 0}^{\infty} f_k$ in an infinite sum of surface spherical harmonics and define
\begin{eqnarray}
\sum\limits_{n = 1}^{\infty}~_kY_n &=& 5 a \left(\boldsymbol{\nabla} \left(f_k\right) \cdot \boldsymbol{E} \cdot \boldsymbol{r} - 3 f_k \frac{\boldsymbol{r} \cdot \boldsymbol{E} \cdot \boldsymbol{r}}{r^2} \right), \label{eq:SB_11b} \\
\sum\limits_{n = 1}^{\infty}~_kZ_n &=& 5 a \left(\boldsymbol{E} \cdot \frac{\boldsymbol{r}}{r}\right)\cdot \boldsymbol{\nabla} \times \left(f_k \boldsymbol{r}\right), \label{eq:SB_11c}
\end{eqnarray}
with the following relations
\begin{eqnarray}
\sum\limits_{k = 0}^{\infty} \sum\limits_{n=1}^{\infty}~_kY_n &=& \sum\limits_{n=1}^{\infty}~Y_n, \label{eq:SB_11d} \\ 
\sum\limits_{k = 0}^{\infty} \sum\limits_{n=1}^{\infty}~_kZ_n &=& \sum\limits_{n=1}^{\infty}~Z_n. \label{eq:SB_11e}
\end{eqnarray}
It can be shown that for every fixed value of $k$, only a finite number of terms in the sums of equations (\ref{eq:SB_11b}) and (\ref{eq:SB_11c}) are non-vanishing. After some algebra, the following relations are obtained
\begin{eqnarray}
_kY_n &=& \left\{
    \begin{array}{ll}
      \frac{5 a k (k-2)}{8 k^2 - 2} \boldsymbol{E} : \left[4 \boldsymbol{\nabla} (f_k) \boldsymbol{r} + 2(k-2) f_k \frac{\boldsymbol{r} \boldsymbol{r} }{r^2}+ \frac{2}{k} r^2 \boldsymbol{\nabla}\boldsymbol{\nabla} (f_k)\right], & n = k-2 \\[2pt]
       \frac{30 a}{8k(k+1)-6} \boldsymbol{E} : \left[\boldsymbol{\nabla} (f_k)  \boldsymbol{r} - k(k+1) f_k \frac{\boldsymbol{r} \boldsymbol{r}}{r^2}  +  r^2 \boldsymbol{\nabla}\boldsymbol{\nabla} (f_k)\right], & n = k \\[2pt]
       \frac{5 a(k+1)(k+3)}{4k(k+2)+3} \boldsymbol{E} : \left[2 \boldsymbol{\nabla} (f_k)  \boldsymbol{r} - (k+3) f_k \frac{\boldsymbol{r} \boldsymbol{r}}{r^2}- \frac{1}{k+1} r^2 \boldsymbol{\nabla}\boldsymbol{\nabla} (f_k)\right], & n = k +2 \\[2pt]
       0, & \mathrm{else},
  \end{array} \right.  \label{eq:SB_13}
\end{eqnarray}
\begin{eqnarray}
_kZ_n &=& \left\{
    \begin{array}{ll}
      \frac{5 a}{2k+1} \left[k \left(\boldsymbol{E} \cdot \frac{\boldsymbol{r}}{r} \right) \cdot \left( \boldsymbol{\nabla}(f_k) \times \boldsymbol{r}\right)+ r \boldsymbol{E} : \boldsymbol{\nabla} \left( \boldsymbol{\nabla} (f_k) \times \boldsymbol{r} \right) \right], & n = k-1 \\[2pt]
       \frac{5 a}{2k+1} \left[(k+1) \left(\boldsymbol{E} \cdot \frac{\boldsymbol{r}}{r} \right) \cdot \left( \boldsymbol{\nabla}(f_k) \times \boldsymbol{r}\right) \right. \\
       \left. - r \boldsymbol{E} : \boldsymbol{\nabla} \left( \boldsymbol{\nabla} (f_k) \times \boldsymbol{r} \right) \right], & n = k+1 \\[2pt]
       0, & \mathrm{else.}
  \end{array} \right. \label{eq:SB_14}
\end{eqnarray}
These identities can be verified in two steps: (i) The sum of all non-zero contributions on the right-hand side of equations (\ref{eq:SB_13}) and (\ref{eq:SB_14}) have to be equal the right-hand sides of equations (\ref{eq:SB_11b}) and (\ref{eq:SB_11c}), (ii) Since $_kY_n$ and $_kZ_n$ are surface spherical harmonics, it follows that $\boldsymbol{\nabla}^2 (r^n \lambda_n) = 0~ \forall n$, with $\lambda_n =~_kY_n,~_kZ_n$. \\
Using the following three solid spherical harmonics
\begin{eqnarray}
_kp_{-(n+1)} &=& \frac{(2 n - 1) \mu}{(n+1) a} \left(\frac{a}{r} \right)^{n+1} ~_kY_n ,\label{eq:SB_15} \\
_k\phi_{-(n+1)} &=& \frac{a}{2(n+1)} \left(\frac{a}{r} \right)^{n+1} ~_kY_n ,\label{eq:SB_16}\\
_k\chi_{-(n+1)} &=& \frac{1}{n(n+1)} \left(\frac{a}{r} \right)^{n+1}~_kZ_n,\label{eq:SB_17}
\end{eqnarray}
the velocity and pressure field can be written as \citep{Brenner1964}
\begin{eqnarray}
\boldsymbol{u} &=& \sum\limits_{n = 1}^{\infty} \left[ \boldsymbol{\nabla} \times \left( \boldsymbol{r} ~_k\chi_{-(n+1)} \right) + \boldsymbol{\nabla} ~_k\phi_{-(n+1)} \right. \nonumber \\
&& \left. - \frac{(n-2)}{2n(2n-1) \mu} r^2 \boldsymbol{\nabla} ~_kp_{-(n+1)} + \boldsymbol{r} \frac{(n+1)}{n(2n-1)\mu} ~_kp_{-(n+1)} \right] \label{eq:SB_18}, \\
p &=& \sum\limits_{n = 1}^{\infty} ~_kp_{-(n+1)}. \label{eq:SB_19}
\end{eqnarray}
Using equations (\ref{eq:SB_13}) and (\ref{eq:SB_14}) we may rewrite equations (\ref{eq:SB_15}) - (\ref{eq:SB_17}) by replacing $n$ with $k$ for all non-vanishing contributions
\begin{eqnarray}
_k p_{-(k-1)} &=& \frac{5(k-2)k(2k-5)}{2(k-1)(4k^2 -1)} \mu \left(\frac{a}{r} \right)^{k-1} \boldsymbol{E}:\left[ 4 \boldsymbol{\nabla} (f_k) \boldsymbol{r} + 2(k-2) f_k \frac{\boldsymbol{r} \boldsymbol{r}}{r^2} \right. \nonumber \\
&& \left. + \frac{2}{k}r^2 \boldsymbol{\nabla} \boldsymbol{\nabla} f_k\right], \label{eq:SB_20} \\
_k p_{-(k+1)} &=& \frac{15}{2k^2+5k+3} \mu  \left(\frac{a}{r} \right)^{k+1} \boldsymbol{E}:\left[ \boldsymbol{\nabla} (f_k) \boldsymbol{r} - k(k+1) f_k \frac{\boldsymbol{r} \boldsymbol{r}}{r^2} \right. \nonumber \\
&& \left. + r^2 \boldsymbol{\nabla} \boldsymbol{\nabla} f_k \right], \label{eq:SB_21} \\
_k p_{-(k+3)} &=& \frac{5(k+1)}{2k+1} \mu \left(\frac{a}{r} \right)^{k+3} \boldsymbol{E}: \left[2 \boldsymbol{\nabla} (f_k) \boldsymbol{r} - (k+3) f_k \frac{\boldsymbol{r} \boldsymbol{r}}{r^2} \right. \nonumber \\
&& \left. - \frac{1}{k+1} r^2 \boldsymbol{\nabla} \boldsymbol{\nabla} f_k \right], \label{eq:SB_22} \\
_k \phi_{-(k-1)} &=& \frac{5k(k-2)a^2}{4(k-1)(4k^2-1)} \left(\frac{a}{r} \right)^{k-1} \boldsymbol{E}: \left[ 4 \boldsymbol{\nabla} (f_k) \boldsymbol{r} + 2(k-2) f_k \frac{\boldsymbol{r} \boldsymbol{r}}{r^2} \right. \nonumber \\
&& \left. + \frac{2}{k}r^2 \boldsymbol{\nabla} \boldsymbol{\nabla} f_k\right], \label{eq:SB_23} \\
_k \phi_{-(k+1)} &=& \frac{15a^2}{(k+1)(8k(k+1)-6)} \left(\frac{a}{r} \right)^{k+1} \boldsymbol{E}:\left[ \boldsymbol{\nabla} (f_k) \boldsymbol{r} - k(k+1) f_k \frac{\boldsymbol{r} \boldsymbol{r}}{r^2} \right. \nonumber \\
&& \left. + r^2 \boldsymbol{\nabla} \boldsymbol{\nabla} f_k \right], \label{eq:SB_24} \\
_k \phi_{-(k+3)} &=& \frac{5(k+1)a^2}{8k(k+2)+6} \left(\frac{a}{r} \right)^{k+3} \boldsymbol{E}:\left[2 \boldsymbol{\nabla} (f_k) \boldsymbol{r} - (k+3) f_k \frac{\boldsymbol{r} \boldsymbol{r}}{r^2} \right. \nonumber \\
&& \left. - \frac{1}{k+1} r^2 \boldsymbol{\nabla} \boldsymbol{\nabla} f_k \right], \label{eq:SB_25} \\
_k \chi_{-k} &=& \frac{5 a}{2k^3 - k^2 - k} \left(\frac{a}{r} \right)^{k}  \left[ k \left(\boldsymbol{E}\cdot \frac{\boldsymbol{r}}{r} \right)\cdot \left(\boldsymbol{\nabla}(f_k) \times \boldsymbol{r} \right) \right. \nonumber \\
&& + \left. r \boldsymbol{E} : \boldsymbol{\nabla} (\boldsymbol{\nabla} (f_k) \times \boldsymbol{r})  \right], \label{eq:SB_26} \\
_k \chi_{-(k+2)} &=& \frac{5 a}{2k^3 + 7k^2 + 7k+2}\left(\frac{a}{r} \right)^{k+2} \left[ (k+1) \left(\boldsymbol{E}\cdot \frac{\boldsymbol{r}}{r} \right)\cdot \left(\boldsymbol{\nabla}(f_k) \times \boldsymbol{r} \right) \right. \nonumber \\
&& \left. - r \boldsymbol{E} : \boldsymbol{\nabla} (\boldsymbol{\nabla} (f_k) \times \boldsymbol{r})  \right]. \label{eq:SB_27}
\end{eqnarray}
The pressure up to first order in $\beta$ is then found via
\begin{eqnarray}
p_0 &=& \left. _k p_{-(k+3)}\right|_{k = 0}, \label{eq:SB_28} \\
p_1 &=& \left. _k p_{-(k+1)}\right|_{k = 1} + \left. _k p_{-(k+3)}\right|_{k = 1}, \label{eq:SB_29} \\
p_2 &=& \left. _k p_{-(k+1)}\right|_{k = 2} + \left. _k p_{-(k+3)}\right|_{k = 2}, \label{eq:SB_30} \\
p_k &=& _k p_{-(k-1)} +~_k p_{-(k+1)} +~_k p_{-(k+3)}, \mathrm{~if~} k>2, \label{eq:SB_31} \\
p^{(1)} &=& \sum\limits_{k = 0}^{\infty} p_k. \label{eq:SB_32}
\end{eqnarray}
Following the same strategy, we find for the velocity field $\boldsymbol{u}$ up to first order in $\beta$:
\begin{eqnarray}
\boldsymbol{u}_{0} &=& \boldsymbol{\nabla} \times \left( \boldsymbol{r} \left. _k\chi_{-(k+2)}\right|_{k = 0} \right) + \boldsymbol{\nabla} \left(\left. _k \phi_{-(k+3)} \right|_{k = 0} \right) + \frac{1}{2 \mu} \boldsymbol{r} \left. _k p_{-(k+3)}\right|_{k = 0}, \label{eq:SB_33} \\
\boldsymbol{u}_{1} &=&  \boldsymbol{\nabla} \times \left( \boldsymbol{r} \left. _k\chi_{-(k+2)}\right|_{k = 1} \right) + \boldsymbol{\nabla} \left(\left. _k \phi_{-(k+3)} \right|_{k = 1} + \left. _k \phi_{-(k+1)} \right|_{k = 1} \right) \nonumber \\
&& + \frac{1}{2 \mu} r^2 \boldsymbol{\nabla} \left( \left. _k p_{-(k+1)} \right|_{k = 1} - \frac{1}{15} \left. _k p_{-(k+3)} \right|_{k = 1}  \right) \nonumber \\
&& + \frac{2}{\mu} \boldsymbol{r} \left( \left. _k p_{-(k+1)}\right|_{k = 1}  + \frac{2}{15} \left. _k p_{-(k+3)} \right|_{k = 1} \right), \label{eq:SB_34} \\
\boldsymbol{u}_{2} &=& \boldsymbol{\nabla} \times \left[ \boldsymbol{r} \left( \left. _k\chi_{-(k+2)}\right|_{k = 2} + \left. _k\chi_{-k}\right|_{k = 2} \right) \right] \nonumber \\
&& + \boldsymbol{\nabla} \left(\left. _k \phi_{-(k-1)} \right|_{k = 2} + \left. _k \phi_{-(k+3)} \right|_{k = 2} + \left. _k \phi_{-(k+1)} \right|_{k = 2}\right) \nonumber \\
&& - \frac{r^2}{28 \mu} \boldsymbol{\nabla} \left( \left. _k \phi_{-(k+3)} \right|_{k = 2} \right) + \frac{1}{2 \mu} \boldsymbol{r} \left( \left. _k p_{-(k+1)}\right|_{k = 2} + \frac{5}{14} \left. _k p_{-(k+3)}\right|_{k = 2} \right), \label{eq:SB_35} \\
\boldsymbol{u}_{k} &=& \boldsymbol{\nabla} \times \left[ \boldsymbol{r} \left( _k\chi_{-(k+2)} +  _k\chi_{-k} \right) \right] + \boldsymbol{\nabla} \left( _k \phi_{-(k-1)}  +  _k \phi_{-(k+3)} + _k \phi_{-(k+1)} \right) \nonumber \\
&& - \frac{(k-4)}{2(k-2)(2(k-2) -1)\mu} r^2 \boldsymbol{\nabla} \left( _k p_{-(k-1)} \right) - \frac{(k-2)}{2k(2k-1) \mu} r^2 \boldsymbol{\nabla} \left( _k p_{-(k+1)} \right)  \nonumber \\
&& - \frac{k}{2(k+2)(2(k+2)-1) \mu} r^2 \boldsymbol{\nabla} \left( _k p_{-(k+3)} \right) \nonumber \\
&& + \boldsymbol{r} \frac{k-1}{(k-2)(2(k-2)-1) \mu} ~_k p_{-(k-1)} + \boldsymbol{r} \frac{(k+1)}{k(2k-1) \mu}~_k p_{-(k+1)} \nonumber \\
&& \boldsymbol{r} \frac{k+3}{(k+2)(2(k+2) -1) \mu} ~_k p_{-(k+3)} \mathrm{,~if~} k>2,\label{eq:SB_36} \\
\boldsymbol{u}^{(1)} &=& \sum\limits_{k = 0}^{\infty} \boldsymbol{u}_{k}. \label{eq:SB_37}
\end{eqnarray}
Equations (\ref{eq:SB_20}) - (\ref{eq:SB_37}) represent the first-order solution of the Stokes equations. The stresslet can be calculated in its most general form for Lambs general solution and reads \citep{Kim2013}
\begin{equation}
\boldsymbol{S} = - \frac{2 \upi}{3} \boldsymbol{\nabla} \boldsymbol{\nabla} \left(r^5 p_{-3}\right). \label{eq:SB_38}
\end{equation}
With the help of the foregoing derivations, we rewrite the $O(\beta)$ contribution to the latter equation as
\begin{eqnarray}
\boldsymbol{S}^{(1)} &=& - \frac{2 \upi}{3} \boldsymbol{\nabla} \boldsymbol{\nabla} \left[r^5 \left(\left. _k p_{-(k-1)} \right|_{k = 4} + \left. _k p_{-(k+1)} \right|_{k = 2} + \left. _k p_{-(k + 3)} \right|_{k = 0} \right)\right]. \label{eq:SB_39}
\end{eqnarray}
When inserting equations (\ref{eq:SB_20}) - (\ref{eq:SB_22}), we find after some algebra
\begin{eqnarray}
\boldsymbol{S}^{(1)} &=& 20 \upi \mu a^3 \left(f_0 \boldsymbol{E} - \frac{1}{21} \left(\boldsymbol{\nabla} \boldsymbol{\nabla} \left(r^2 f_2 \right) : \boldsymbol{E} \right) \boldsymbol{I} \right. \nonumber \\
&&  \left. + \frac{1}{14} \left(\boldsymbol{E} \cdot \boldsymbol{\nabla} \boldsymbol{\nabla} \left(r^2 f_2 \right) + \boldsymbol{\nabla} \boldsymbol{\nabla} \left(r^2 f_2 \right) \cdot \boldsymbol{E}   \right) \right. \nonumber \\
&& \left. - \frac{1}{189} \boldsymbol{\nabla} \boldsymbol{\nabla} \boldsymbol{\nabla} \boldsymbol{\nabla} \left(r^4 f_4 \right) : \boldsymbol{E} \right). \label{eq:SB_40}
\end{eqnarray}
It should be noted that the latter equation holds for an arbitrary slightly deformed sphere. Interestingly, compared to the force and torque acting on a slightly deformed sphere \citep{Brenner1964} which are determined by $f_0$, $f_1$ and $f_2$, the stresslet requires more detailed information of the shape of the particle, reflected in the term containing  $f_4$. 

{\color{black}\section{Details on the solution of the Stokes equations around a mirror-fused particle in a pure straining flow} \label{sec:AppEff}
\subsection{Zeroth-order solution}\label{sec:Eff_1}
In this section we compute the effective surface viscosities for the case $\alpha = 90^\circ$. This special case has already been studied by \cite{Lishchuk2009} and can therefore be considered as a benchmark for our approach. The stresslet of a spherical particle in a pure straining flow is readily known \citep[e.g.,][]{Kim2013} and therefore the double contraction with the rate of strain tensor form equation (\ref{eq:Prob_6}) can be evaluated, leading to
\begin{equation}
\boldsymbol{S}^{(0)} : \boldsymbol{E} = \frac{20 \upi}{3} \mu a^3 \left[\left(\langle \boldsymbol{\nabla}^{\mathcal{S}} \cdot \boldsymbol{u}^{\mathcal{S}} \rangle\right)^2 + \left(\langle \boldsymbol{E}^\mathcal{S} \rangle\right)^2 \right], \label{eq:Eff_16}
\end{equation} 
from which the constants in equation (\ref{eq:Eff_9}) can be read off:
\begin{eqnarray}
C_1^{(0)} = C_2^{(0)} = \frac{20}{3}. \label{eq:Eff_17}
\end{eqnarray}
The zeroth-order interfacial viscosities follow from equations (\ref{eq:Eff_14}) \& (\ref{eq:Eff_15})
\begin{eqnarray}
\mu^{\mathcal{S}(0)} &=& \frac{5}{3} \phi \mu a \label{eq:Eff_18}, \\
\kappa^{\mathcal{S}(0)} &=& 5 \phi \mu a \label{eq:Eff_19}.
\end{eqnarray}
The zeroth-order solutions are identical to the results reported in \cite{Lishchuk2009} and therefore provide verification for the method presented in section {\color{black}\ref{sec:Prob_1}}.\\

Before we proceed to calculate the first-order corrections of the interfacial viscosities, we introduce the Lorentz reciprocal theorem \citep{Lorentz1896a} as a useful tool for the following analysis.  

\subsection{Lorentz reciprocal theorem}\label{sec:Eff_2}
In its most general form, the Lorentz reciprocal theorem reads \citep{Lorentz1896a,Brenner1991,Kim2013}
\begin{equation}
\int\limits_{A_\mathcal{P}} \boldsymbol{u} \cdot \bar{\boldsymbol{\Pi}} \cdot \boldsymbol{n}~\mathrm{d }A = \int\limits_{A_\mathcal{P}} \bar{\boldsymbol{u}} \cdot \boldsymbol{\Pi} \cdot  \boldsymbol{n}~ \mathrm{d }A, \label{eq:Eff_20}
\end{equation}
in which ($\boldsymbol{u}$, $\boldsymbol{\Pi}$) and ($\bar{\boldsymbol{u}}$, $\bar{\boldsymbol{\Pi}}$) satisfy the Stokes equations for the same domain but with different boundary conditions. In the present context, ($\bar{\boldsymbol{u}}$, $\bar{\boldsymbol{\Pi}}$) correspond to a spherical particle with the boundary condition (\ref{eq:Eff_3}) and a vanishing velocity field at $r \to \infty$. The surface stress for this case reads $\bar{\boldsymbol{\Pi}} \cdot \boldsymbol{n} = 3 \mu a\boldsymbol{E} \cdot \frac{\boldsymbol{r}}{r}$ \citep{Happel2012}. Inserting the expression for the surface stress as well as equation (\ref{eq:Eff_3}) (here: $\tilde{\boldsymbol{u}} = - \boldsymbol{E} \cdot \boldsymbol{r} a/r$) into equation (\ref{eq:Eff_20}), we get 
\begin{equation}
\left(3 \mu~ \int\limits_{A_\mathcal{P}} \boldsymbol{u} \boldsymbol{n}~\mathrm{d }A\right) : \boldsymbol{E} = \left(- \int\limits_{A_\mathcal{P}} \boldsymbol{r}  \boldsymbol{\Pi} \cdot \boldsymbol{n} ~\mathrm{d }A\right):\boldsymbol{E}. \label{eq:Eff_21}
\end{equation}
Since this equation needs to hold independent of the choice of $\boldsymbol{E}$, the two terms inside the brackets need to be identical. When calculating the symmetric and traceless part of the terms inside the brackets, we find 
\begin{eqnarray}
\int\limits_{A_\mathcal{P}} \frac{3 \mu}{2} \left(\boldsymbol{u} \boldsymbol{n} + \boldsymbol{n} \boldsymbol{u} \right) - \mu (\boldsymbol{u} \cdot \boldsymbol{n}) \boldsymbol{I}~\mathrm{d }A &=& - \int\limits_{A_\mathcal{P}} \frac{1}{2} \left(\boldsymbol{r} \boldsymbol{\Pi} + \boldsymbol{\Pi} \boldsymbol{r} \right) \cdot \boldsymbol{n} \nonumber \\
&& - \frac{1}{3} (\boldsymbol{r} \cdot \boldsymbol{\Pi} \cdot \boldsymbol{n})~ \boldsymbol{I}  ~\mathrm{d }A \label{eq:Eff_22}.
\end{eqnarray}
For a rigid particle we have $\boldsymbol{u} \cdot \boldsymbol{n} = 0$. When adding $\int_{A_\mathcal{P}} \mu (\boldsymbol{u} \boldsymbol{n} + \boldsymbol{n} \boldsymbol{u}) \mathrm{d }A$ to both sides of the latter equation, the right-hand side of equation (\ref{eq:Eff_22}) corresponds to the negative stresslet. Therefore, we finally find
\begin{equation}
\boldsymbol{S} =  - \frac{5 \mu}{2}~\int\limits_{A_\mathcal{P}} ( \boldsymbol{u} \boldsymbol{n} + \boldsymbol{n} \boldsymbol{u} ) ~ \mathrm{d }A  \label{eq:Eff_23}.
\end{equation} 
In contrast to equation (\ref{eq:Prob_11}), in which the calculation of the surface stress requires detailed information on the velocity and pressure fields around the particle, equation (\ref{eq:Eff_23}) only involves an integration of the boundary conditions over the particle surface. It should be noted that the validity of equation (\ref{eq:Eff_23}) can be verified with the help of Lamb's general solution \citep{Happel2012, Kim2013}. After a lengthy calculation the expression of equation (\ref{eq:SB_38}) is recovered.

\subsection{First-order solution}\label{sec:Eff_3}
Equations (\ref{eq:Eff_3}) - (\ref{eq:Eff_5}) display the successive solution process for slightly deformed spheres, i.e. the $n$-th order boundary condition is solely determined by the solutions up to $(n-1)$-th order. Therefore, the first-order correction of the stresslet acting on a slightly deformed sphere can be calculated by inserting equation (\ref{eq:Eff_4}) in equation (\ref{eq:Eff_23}). Alternatively, the stresslet can be calculated according to equation (\ref{eq:SB_38}), but requires the solution of the Stokes equations to first order in $\beta$. In appendix \ref{sec:S2} we derive the first-order solution for a slightly deformed sphere in pure straining and provide an expression for the stresslet [see equation (\ref{eq:SB_40})]. Making use of these results, we find after a short calculation 
\begin{eqnarray}
\boldsymbol{S}^{(1)}: \boldsymbol{E} = 20 \upi \mu a^3 \left[\frac{35}{48} \left(\langle \boldsymbol{\nabla}^{\mathcal{S}} \cdot \boldsymbol{u}^{\mathcal{S}} \rangle\right)^2 + \frac{5}{12} \left( \langle \boldsymbol{E}^\mathcal{S} \rangle\right)^2  \right], \label{eq:Eff_24}
\end{eqnarray}
from which the following coefficients can be read off
\begin{eqnarray}
C_1^{(1)} =  \frac{175}{12}, \qquad C_2^{(1)} = \frac{25}{3}. \label{eq:Eff_25}
\end{eqnarray}
The first-order {\color{black} corrections to} effective surface viscosities then read
\begin{eqnarray}
\mu^{\mathcal{S}(1)} &=& \frac{25}{12} \phi \mu a \label{eq:Eff_26}, \\
\kappa^{\mathcal{S}(1)} &=& \frac{75}{8} \phi \mu a \label{eq:Eff_27}.
\end{eqnarray}
It should be noted that, up to first order in $\beta$, both definitions of the surface concentration [equation (\ref{eq:Prob_1}) and (\ref{eq:Prob_2})] lead to the same effective viscosities. 

\subsection{Second-order solution}\label{sec:Eff_4}
To obtain the second-order solution for the stresslet, we make use of the Lorentz reciprocal theorem from section \ref{sec:Eff_2}. The second-order boundary condition at the surface of an undeformed sphere is given in equation (\ref{eq:Eff_5}). For reasons that will become clear later, we decompose equation (\ref{eq:Eff_5}) into two contributions, a first one proportional to the normal gradient of $\boldsymbol{u}^{(1)}$, and a second one containing all other terms. When employing the reciprocal theorem [equation (\ref{eq:Eff_23})], we start by evaluating the integral over all contributions that are not proportional to the normal gradient of $\boldsymbol{u}^{(1)}$ and find
\begin{eqnarray}
\boldsymbol{S}^{(2)}:\boldsymbol{E} &=& - \frac{20 \upi}{7} \mu a^3 \left[ 4 \left(\langle \boldsymbol{\nabla}^{\mathcal{S}} \cdot \boldsymbol{u}^{\mathcal{S}} \rangle\right)^2 + 5 \left( \langle \boldsymbol{E}^\mathcal{S} \rangle\right)^2   \right] \nonumber \\
&& + \frac{5}{2} \mu a \left[\int\limits_{A_\mathcal{P}} f(\theta, \varphi) \left[\left. \frac{\partial \boldsymbol{u}^{(1)}}{\partial r}\right|_a \frac{\boldsymbol{r}}{r} + \frac{\boldsymbol{r}}{r} \left. \frac{\partial \boldsymbol{u}^{(1)}}{\partial r}\right|_a \right] \mathrm{~d}A \right] : \boldsymbol{E}. \label{eq:Eff_28}
\end{eqnarray}
\begin{figure}
  \centerline{\includegraphics[width=0.95\textwidth]{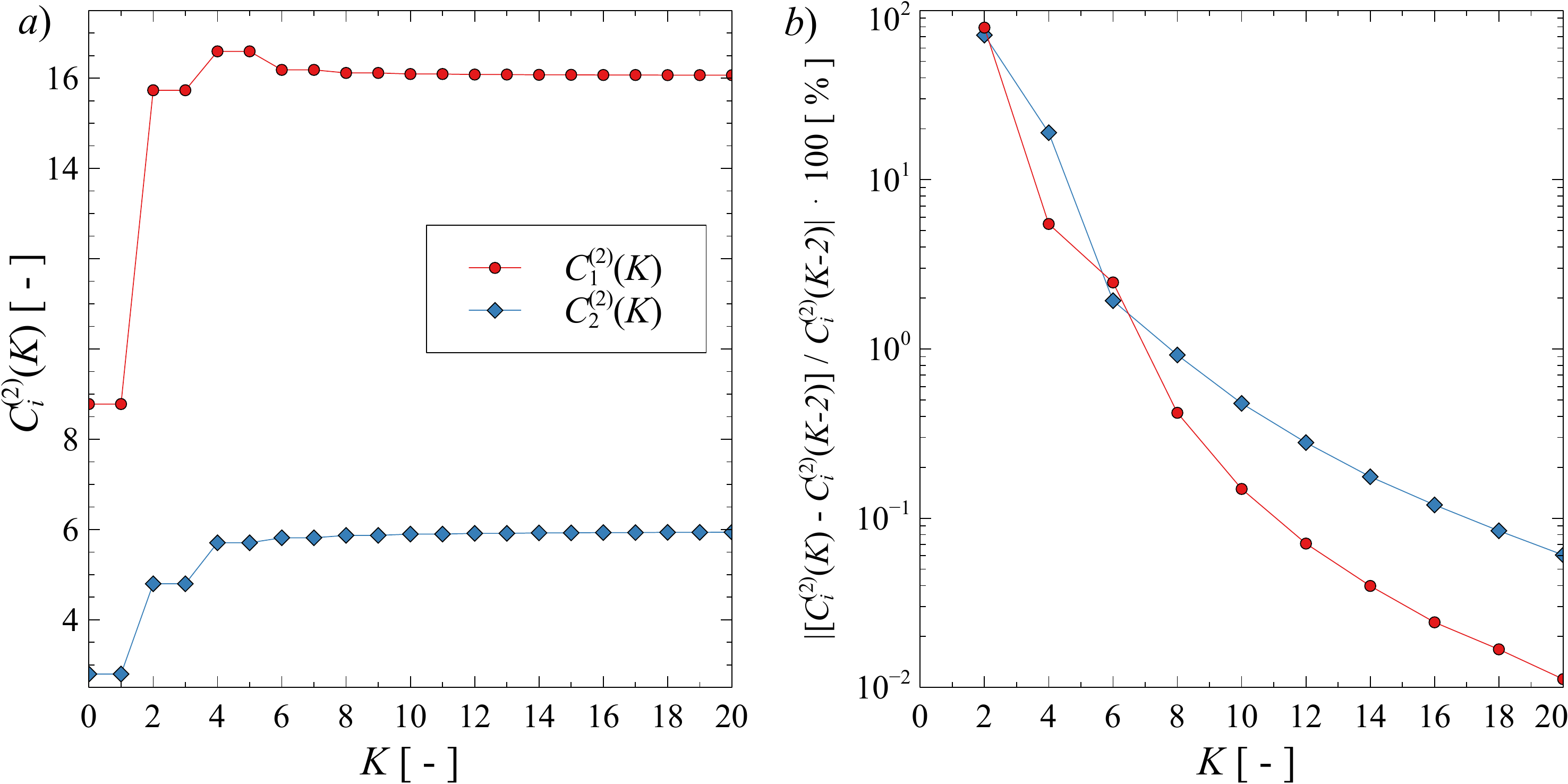}}
  \caption{Convergence plot showing the dependency of the coefficients $C^{(2)}_i (~i = 1,2)$ referring to the second-order calculation of the stresslet of a slightly deformed sphere, on the summation limit $K$. $a$) Absolute value of $C^{(2)}_i (~i = 1,2)$ versus summation limit $K$. $b$) Relative change of $C^{(2)}_i (~i = 1,2)$ between $K+2$ and $K$ in percent. }
\label{fig:f10}
\end{figure}
$\boldsymbol{u}^{(1)}$ is written as an expansion in spherical harmonics [see equation (\ref{eq:SB_37})], reflecting the angular dependence of $f(\theta,\varphi)$. We truncate the infinite sum by setting
\begin{eqnarray}
\boldsymbol{u}^{(1)} \approx \sum\limits_{k = 0}^{K} \boldsymbol{u}^{(1)}_k. \label{eq:Eff_29}
\end{eqnarray}
Inserting equation (\ref{eq:Eff_29}) into equation (\ref{eq:Eff_28}), and evaluating the remaining integral in equation (\ref{eq:Eff_28}) for $K = 20$, we find
\begin{eqnarray}
\boldsymbol{S}^{(2)}:\boldsymbol{E} &\approx & \upi \mu a^3 \left[ \frac{1555263045342049}{165476499980288} \left(\langle \boldsymbol{\nabla}^{\mathcal{S}} \cdot \boldsymbol{u}^{\mathcal{S}} \rangle\right)^2 \right. \nonumber \\
&& \left. - \frac{60053874226845}{82738249990144}\left( \langle \boldsymbol{E}^\mathcal{S} \rangle\right)^2  \right]. \label{eq:Eff_30}
\end{eqnarray}
With the help of equations (\ref{eq:Eff_9}) \& (\ref{eq:Eff_11}), the second-order coefficients are obtained as
\begin{eqnarray}
C_1^{(2)} \approx \frac{7975319135631907}{496429499940864} \approx 16.065, \quad C_2^{(2)} \approx \frac{1474603377122345}{248214749970432} \approx 5.941 \label{eq:Eff_31}.
\end{eqnarray}
To show that the 20 partial sums are sufficient to approximate the velocity field around the particle reasonable well, we plot the constants $C_1^{(2)}$ and $C_2^{(2)}$ as a function of $K$ in figure \ref{fig:f10}. Figure \ref{fig:f10} $a$) shows both coefficients as a function of $K$. The relative error of both coefficients decreases with increasing $K$ and falls below 0.1 \% for $K = 20$, as shown in figure \ref{fig:f10} $b$). All calculated coefficients are collected in table \ref{tab:t1}. The second-order correction of the surface viscosities are computed for two different definitions of the surface concentration, i.e., equation (\ref{eq:Prob_1}) and (\ref{eq:Prob_2}), as follows
\begin{eqnarray}
\mu^{\mathcal{S}(2)} &\approx & 1.4853 \phi \mu a, \quad \mathrm{ or } \quad \mu^{\mathcal{S}(2)} \approx  -0.1815 \phi_L \mu a \label{eq:Eff_32}\\
\kappa^{\mathcal{S}(2)} &\approx & 9.5178 \phi \mu a, \quad \mathrm{ or } \quad \kappa^{\mathcal{S}(2)} \approx 4.5179 \phi_L \mu a. \label{eq:Eff_33}
\end{eqnarray}}

\section{Calculation of the effective surface tension $\gamma^\mathcal{S}$}\label{sec:S3}
As already mentioned in section \ref{sec:NumSim}, the effective surface tension is defined as the ratio of the surface energy $E_\mathrm{SF}$ and the area of the interface $A$. The energy of a particle-laden interface can be calculated via
\begin{equation}
E_\mathrm{SF} = \gamma_{12} A_{12} + N \left( \gamma_{\mathcal{P}(1)} A_{\mathcal{P}(1)} + \gamma_{\mathcal{P}(2)} A_{\mathcal{P}(2)} \right),\label{eq:SC_1}
\end{equation}
in which $\gamma_{12}$, $A_{12}$, $\gamma_{\mathcal{P}(i)}$, $A_{\mathcal{P}(i)}$ and $N$ is the interfacial tension between the two fluids, the area of the interface between the two fluids, the interfacial tension between a particle and phase $i$, the interfacial area between a particle and phase $i$, and the number of interfacial particles, respectively. The total area of the interface is $A = A_{12} + N \upi a^2 (1- \cos^2(\alpha))$, or when using equation (\ref{eq:Prob_2})
\begin{equation}
A_{12} = A (1- \phi).\label{eq:SC_2}
\end{equation}   
The solid-fluid interfacial areas of a particle are
\begin{eqnarray}
A_{\mathcal{P}(1)} &=& 2 \upi a^2 (1+ \cos(\alpha)) = 2 \frac{1+\cos(\alpha)}{1 - \cos^2(\alpha)}  \frac{\phi A}{N} \approx 2 \left[1 + \cos(\alpha) + \cos^2(\alpha)\right] \frac{\phi A}{N},\label{eq:SC_3} \\
A_{\mathcal{P}(2)} &=& 2 \upi a^2 (1- \cos(\alpha)) \approx 2 \left[1 - \cos(\alpha) + \cos^2(\alpha)\right] \frac{\phi A}{N}.\label{eq:SC_4}
\end{eqnarray}
In these equations, an expansion up to $O(\cos^2(\alpha))$ was employed. When inserting equations (\ref{eq:SC_2}), (\ref{eq:SC_3}) \& (\ref{eq:SC_4}) into equation (\ref{eq:SC_1}) and rearranging the terms, we find
 \begin{equation}
E_\mathrm{SF} = \left[\gamma_{12} \left(1 - \phi \right) + 2 (1 + \cos^2(\alpha)) \phi \left(\gamma_{\mathcal{P}(1)}+\gamma_{\mathcal{P}(2)} \right) + 2 \cos(\alpha) \phi \left(\gamma_{\mathcal{P}(1)} - \gamma_{\mathcal{P}(2)} \right) \right] A. \label{eq:SC_5}
\end{equation}
The term in square brackets represents the effective surface tension $\gamma^\mathcal{S}$. After employing Young's law, i.e., $\cos(\alpha) = (\gamma_{\mathcal{P}(2)} -\gamma_{\mathcal{P}(1)})/\gamma_{12}$, equation (\ref{eq:Diss_16}) is obtained.

\bibliographystyle{jfm}
\bibliography{manuscript}

\end{document}